\documentclass[11pt]{article}

\usepackage{color}

\usepackage{epsfig}

\usepackage{epstopdf}
\usepackage{amsmath}
\usepackage{amssymb}
\usepackage{mathrsfs}
\usepackage{graphicx}
\usepackage{float}
\usepackage{subcaption}

\usepackage[colorlinks = true,
linkcolor = magenta,
urlcolor  = blue,
citecolor = red,
anchorcolor = blue]{hyperref}

\begin{document}
	\def \a'{\alpha'}
	\baselineskip 0.65 cm
	\begin{flushright}
		\ \today
	\end{flushright}

\begin{center}{\large
		{\bf A Minimal Model For Two-Component FIMP Dark Matter: A Basic Search }} {\vskip 0.5 cm} {\bf ${\rm S.~ Peyman~ Zakeri}$$^1$, ${\rm S.~Mohammad~Moosavi~Nejad}$$^{1,2}$, ${\rm  Mohammadreza~ Zakeri}$$^{3,4}$ and ${\rm S.~Yaser~Ayazi}$$^5$}{\vskip 0.5 cm
	}
	{\small $^1$$Faculty~of~Physics$,~ $Yazd~University$,~$P.O.~Box~89195-741$,~$Yazd$,~$Iran$}{\vskip 0.1 cm}{\bf \small 
		$^2$$School~ of~ Particles~ and~ Accelerators,~ Institute~ for~ Research~ in~
		Fundamental$ $ Sciences~(IPM), P.O.~Box~ 19395-5531, Tehran, Iran$
		$^3$$Physics~and~Astronomy~Department$,~$ University~of~California$, $Riverside$,~$California~92521$, $USA$
		$^4$$CAS~Key~Laboratory~of~Theoretical~Physics$,~$ Institute~of~Theoretical~Physics$, $Chinese~Academy~of~Sciences$,~$Zhong~Guan~Cun~East~Rd.55$,~$Beijing, 100190$,~$China$
		$^5$$Physics~Department$,~		$Semnan~University$,~$P.O.~Box~35131-19111$,~$Semnan$,~$Iran$}

\end{center}

\begin{abstract}
	
In the multi-component configurations of dark matter phenomenology, we propose a minimal two-component configuration which is an extension of the Standard Model with only three new fields; one scalar and one fermion interact with the thermal soup through Higgs portal, mediated by  the other scalar in such a way that the stabilities of dark matter candidates are made simultaneously by an explicit $Z_2$ symmetry. Against the most common freeze-out framework, we look for dark matter particle signatures in the freeze-in scenario by evaluating the relic density and detection signals. A simple distinguishing feature of the model is the lack of dark matter conversion, so the dark matter components act individually and the model can be adapted entirely to both singlet scalar and singlet fermionic models, separately. We find dark matter self-interaction as the most promising approach to probe such feeble models. Although the scalar component satisfies this constraint, the fermionic one refuses it even in the resonant region.  
      
\end{abstract}

%\pacs{14.65.Ha, 13.88.+e, 14.40.Lb, 14.40.Nd}

%\maketitle
\section{Introduction} 
\label{sec1}

Weakly interacting massive particles (WIMPs) are the most popular solution to the puzzle of dark matter (DM) \cite{Gondolo:1990dk,Srednickii,Chiu}. In TeV scale (LHC scale) new physics, DM particles follow the thermal scenario in which they reach thermal and chemical equilibrium with the bath particles but lose it at the freeze-out temperature (which is around $m_{\text{DM}}/20$) and experience decoupling from the Universe plasma. WIMP candidates such as the neutralino \cite{Jungman:1995df}  and Kaluza-Klein particle \cite{Cheng:2002ej,Servant:2002aq} are found in theories such as the minimal supersymmetric Standard Model (MSSM) and universal extra dimensions (UED), respectively, and also in other extensions of the Standard Model (SM) such as singlet scalar \cite{Silveira:1985rk,McDonald,Burgess,Barger} (or fermionic \cite{Kim:2008pp,Ettefaghi,Fairbairn}) DM. In spite of their popularity,  WIMPs have not yet been  detected in  direct experiments. 

The other viable and well-motivated hypothesis to explain the DM problem is that there is such a feeble interaction that DM particles can never be abundant enough to thermalize. In this so-called freeze-in mechanism  \cite{McDonald:2001vt, Hall:2009bx,Bernal:2017kxu}, feebly interacting massive particles (FIMPs)  have been  slowly produced in the early Universe through the collisions or decays of the bath particles. FIMP candidates are motivated in various extensions of the SM \cite{Yaguna:2011qn,Klasen:2013ypa,Ayazi:2015jij} and a well-known example which arises from  neutrino physics is the sterile neutrino \cite{Merle:2014xpa,Merle,Shakya,Kang:2014cia,Biswas}.  It is difficult to detect FIMP particles because of their small couplings with the SM. As for the indirect  searches, depending on the type of DM candidate, i.e. scalar \cite{Pandey:2017quk}, fermion \cite{Ayazi:2015jij}, etc, some experiments have parameter space where they could survive but these are very borderline. For a study of the non-thermal properties of dark matter see Ref.~\cite{Dev:2013yza}.

Although a lot of attention has been dedicated to single-particle DM models, some studies have considered DM models with the contribution of more particles in the observed DM density (multi-component DM  \cite{Profumo:2009tb,Gelm1,Gelm2,Dudaa,Herrero-Garcia:2017vrl}). The simplest and the most common case is the union of the singlet scalar (fermionic) and the singlet fermionic (scalar) models which are employed in both freeze-in \cite{Pandey:2017quk,Biswas:2015sva} and freeze-out \cite{Esch:2014jpa, Bhattacharya:2013hva, Biswas:2013nn, Chialva:2012rq} solutions (or intermediate cases \cite{DuttaBanik:2016jzv}). Nevertheless, it remains a mystery whether DM is a single particle or multi-component.

In this paper, we analyze whether the freeze-in approach can
be properly used to produce the observed DM density in our Universe.  We choose a minimal two-component DM model in such a way that both of the components are FIMP particles. Following our hypothesis, we consider a singlet scalar and a Dirac fermion  where   an accidental symmetry guarantees their stabilities and a Higgs portal enables them to interact with the SM particles. The most striking feature of  our model is its simplicity,  as  the two candidates of DM particles do not  couple with each other and the model has separate overlaps with both the singlet scalar model \cite{Yaguna:2011qn} and the singlet fermionic model \cite{Klasen:2013ypa}.  In our work, all contributing processes to the relic density are assumed and supplementary phenomenological aspects  are also included.

Some promising possible signatures of FIMPs which are found to be most reliable in previous works are the $\gamma$-ray excess  observed from the Galactic center (GC) \cite{Biswas:2015sva, DuttaBanik:2016jzv}, the X-ray line at $3.55$~keV \cite{Biswas:2015sva, DuttaBanik:2016jzv}, and DM self-interaction \cite{Pandey:2017quk, DuttaBanik:2016jzv}. To generate the gamma ray excess, the fermionic component should have a pseudoscalar coupling to the mediator in the freeze-out regime \cite{DuttaBanik:2016jzv}. The scalar component which does also couple to the SM Higgs directly \cite{Biswas:2015sva}, should not feature large valued couplings. An X-ray signal with $E_\gamma=3.55$~keV from the XMM-Newton telescope and a similar signal at $3.52$~keV from the Andromeda galaxy (M31) and Perseus Cluster could all  be interpreted by the decay \cite{Modak:2014vva} or the annihilation \cite{Babu:2014pxa} of DM.  However, this requires a definite decay rate and annihilation cross section which is out of reach for our scenario. Therefore, we continue our probe relying only on the DM self-interaction. This non-gravitational interaction is a well-motivated indirect search  as it solves the tensions between observations and simulations of the small-scale structure of DM.    

Following the aforementioned setup, our paper is organized as follows. After introducing the construction of our model and identifying its parameter space in Section~\ref{sec2}, we solve two independent Boltzmann equations in the following section (Section~\ref{sec3}), in order to reach the observed relic density measured by the WMAP and Planck experiments \cite{Ade:2015dga}. In Section~\ref{sec4}, we study the phenomenological implications for both direct and indirect experiments, and summarize our results in Section~\ref{sec5}.                                             

\section{Two-Component FIMP DM}
\label{sec2}

Beyond the SM, we employ three new fields to furnish our model: two scalars ($\chi$ and $S$) and one Dirac fermion ($\psi$), which are all  assumed to be singlet under the SM gauge groups. A discrete $Z_2$ symmetry is applied such that it reads the SM fields and  the $S$-scalar even, and the other two fields  ($\chi$ and $\psi$) odd. This symmetry guarantees the stability of both odd particles in a way that we do not have any terms involving both fields $\psi$ and $\chi$. In this way, the decays of odd particles to  one another  are prevented. Therefore, we can have two DM candidates in our setup by an accidental symmetry. \\
The framework of  our model is constructed by:
\begin{align} \nonumber
\cal L &\supset\dfrac{1}{2}(\partial_{\mu}S)^2+\dfrac{1}{2} (\partial_{\mu}\chi)^2+i\bar{\psi}{\not}\partial\psi\\ \nonumber
&-m_\psi\bar{\psi}\psi-g_sS\bar{\psi}\psi-g_pS\bar{\psi}\gamma^5\psi \\ 
&-V(H,S,\chi), 
\end{align}
where we introduced  the scalar and pseudoscalar interactions with the couplings $g_s$ and $g_p$, respectively, and inserted the scalar interactions in the term  $V(H,S,\chi)$ as     
\begin{align} \nonumber \label{eq:2}
V(H,S,\chi) &=-\mu_H^2H^{\dagger}H+\lambda_H(H^{\dagger}H)^2\\\nonumber
&+\mu_1S+\dfrac{1}{2}\mu_S^2 S^2+\dfrac{1}{3!}\alpha_s S^3+\dfrac{1}{4!}\lambda_S S^4+\dfrac{1}{2}m^2_{0\chi}\chi^2+\dfrac{1}{4!}\lambda_{\chi} \chi^4\nonumber\\
&+\lambda_1S H^{\dagger}H+ \lambda_2S^2 H^{\dagger}H +\lambda_{\chi H} \chi^2H^{\dagger}H+\lambda_3S\chi^2+\lambda_4S^2\chi^2.  
\end{align}
After spontaneous symmetry breaking, the $SU(2)$ Higgs doublet is parametrized as   
\begin{equation}
H = \frac{1}{\sqrt{2}} \left( \begin{array}{c}
0  \\
v_{H}+h
\end{array} \right)\,,
\end{equation}  
where $v_H=246$~GeV, but for the mediator we assume that it does not acquire a vacuum expectation value, i.e. $<S>=0$,  which minimalizes our model too. Now, due to the interaction terms in Eq.~(\ref{eq:2}), $h$ and $S$ mix with each other and form a mass matrix  with the following eigenstates
\begin{align} \nonumber\label{h1h2}
h_1 &= S\sin \theta + h\cos \theta,\\
h_2 &= S\cos \theta - h\sin \theta,
\end{align}
and  the eigenvalues as
\begin{equation}\label{mass}
m^{2}_{h_1,h_2} = \frac{m^{2}_h+m^{2}_S}{2}\pm \frac{m^{2}_h-m^{2}_S}{2} \sqrt{1+y^2},  ~~~ \text{with}~~ y= \frac{2m^{2}_{h,S}}{m^{2}_h-m^{2}_S},
\end{equation}
where $\theta$ is the mixing angle between $h_1$ and $h_2$ such that
\begin{equation}
\tan \theta = \frac{y}{1+\sqrt{1+y^2}}.
\end{equation}
According to the definition of the mixing angle $\theta$, $h_1$ can be considered as the SM-like Higgs observed at the LHC with a mass of about $125$~GeV.  In Eq.~\eqref{mass}, $m_h=\sqrt{2\lambda_H}v_H$, $m_S=(\lambda_2v_H^2+\mu_S^2)^{1/2}$ and $m_{h,S}=\sqrt{\lambda_1v_H}$.

Concerning our parameters, vacuum stability implies that the scalar potential in Eq.~\eqref{eq:2} must be bounded from below. On the other hand, perturbativity does not allow the model parameters to be too large. Eventually, these theoretical conditions can be satisfied  if one has
\begin{align} \label{eq:7}
-2\pi/3&<\lambda_S, \lambda_{\chi}<2\pi/3,\nonumber\\
-4\pi&<\lambda_2, \lambda_{\chi H}<4\pi,\nonumber\\
-8\pi&<\lambda_4, g_s, g_p<8\pi,\nonumber\\
\lambda_2&+\sqrt{\lambda_H\lambda_S}>0,\nonumber\\
\lambda_{\chi H}&+\sqrt{\lambda_H\lambda_{\chi}}>0,\nonumber\\
2\lambda_4&+\sqrt{\lambda_S\lambda_{\chi}}>0,
\end{align}
and
\begin{align} \label{eq:8}
&\Big(\sqrt{2(\lambda_2+\sqrt{\lambda_H\lambda_S})(\lambda_{\chi H}+\sqrt{\lambda_H\lambda_{\chi}})(2\lambda_4+\sqrt{\lambda_S\lambda_{\chi}})}\nonumber\\
+&\sqrt{\lambda_H\lambda_S\lambda_{\chi}}+\lambda_2\sqrt{\lambda_{\chi}}+\lambda_{\chi H}\sqrt{\lambda_S}+2\lambda_4\sqrt{\lambda_H}\Big)>0,
\end{align}
where $\lambda_H$ is the quartic coupling of $H$. Extending the SM with the new fields $\psi$, $\chi$ and $S$ embeds 19 parameters in addition to the SM ones.  They are  $m_S,\ m_\chi, m_\psi, g_s, g_p, \mu_1, \alpha_S, \lambda_S, \lambda_\chi, \lambda_1, \lambda_2, \lambda_{\chi H}, \lambda_3, \lambda_4, m_{h_1}, m_{h_2}$, $\sin\theta, v_H $, and $m_h$.  However, due to the 8 model constraints, 11 independent parameters,  
\begin{equation}
g_s, g_p, \lambda_3, \lambda_4, \lambda_{\chi H}, m_{\psi}, m_{\chi}, m_{h_2},  \sin\theta, 
\end{equation}
remain for the relic abundance and for indirect searches, $\lambda_H$ and $\lambda_{\chi}$ are required. Here, we take a moment to describe the eight constraints which appear in our work. Note that, after spontaneous symmetry breaking the scalar potential given in Eq.~\eqref{eq:2} reads as $V(h,S,\chi)$. 
Therefore, it can be deduced from the potential that:
\begin{eqnarray}
1)&& \quad \dfrac{\partial V}{\partial h}\bigg|_{h=S=\chi=0}=0 \Rightarrow\mu_H^2=\lambda_Hv_H^2,\\
2)&& \quad \dfrac{\partial V}{\partial S}\bigg|_{h=S=\chi=0}=0 \Rightarrow\mu_1=-\dfrac{\lambda_1v_H^2}{2},\\
3)&&\quad m_h^2=-\mu_H^2+3\lambda_Hv_H^2=2\lambda_Hv_H^2,\\
4)&&\quad  m_S^2=\mu_S^2+\lambda_2v_H^2,\\
5)&& \quad m_\chi^2=m^2_{0\chi}+\lambda_{\chi H}v_H^2.
\end{eqnarray} 
Also, the mixing between $S$ and $h$ produces the scalars $h_1$ and $h_2$ so one can conclude that 
\begin{eqnarray}
\dfrac{1}{2}m_h^2h^2+\dfrac{1}{2}m_S^2S^2+\lambda_1v_HSh=\dfrac{1}{2}m_{h_1}^2h_1^2+\dfrac{1}{2}m_{h_2}^2h_2^2.
\end{eqnarray}
Substituting $h_1=S\sin \theta + h\cos \theta$ and $h_2=S\cos \theta - h\sin \theta$ \eqref{h1h2} and using the constraints (3)-(5), one obtains:
\begin{eqnarray}
6)&& \lambda_H=\dfrac{m_{h_1}^2\cos^2\theta+m_{h_2}^2\sin^2\theta}{2v_H^2},\\
7)&& \lambda_2=\dfrac{m_{h_1}^2\sin^2\theta+m_{h_2}^2\cos^2\theta-\mu_S^2}{v_H^2},\\
8)&& \lambda_1=\dfrac{m_{h_1}^2-m_{h_2}^2}{2v_H}\sin2\theta.
\end{eqnarray}
These 8 constraints reduce the 19 free parameters in the model to the 11 independent parameters.
Also, the couplings $\alpha_s$ and $\lambda_2$ can be taken as zero without any ambiguities. However, we consider $\alpha_s\neq 0$ for  future applications.  In the following, we will probe our model parameter space with experimental constraints coming from the relic density, direct and indirect detections.

\section{DM Density}
\label{sec3}

\subsection{Boltzmann Equation}
\label{sec3.1}

Since our model contains two DM candidates, its relic density has contributions from both fields $\psi$ and $\chi$. Therefore, we have to solve two Boltzmann equations for particles which will not reach equilibrium in the freeze-in mechanism where we follow the solution  in Ref.~\cite{Ayazi:2015jij} (following Ref.~\cite{Hall:2009bx}). The  time evolution of number density, $dn_\text{DM}/dt$, for the fermionic DM  is given by  
\begin{align} \nonumber
\label{eq:10}
\dfrac{dn_{\psi}}{dt}+3Hn_{\psi}&= \dfrac{T}{\pi^2}\sum_{i=1}^{2}m_{h_i}^2K_1(\dfrac{m_{h_i}}{T}) \Gamma_{h_i\rightarrow\bar{\psi}\psi} \\
&+\dfrac{T}{32\pi^4}\sum_{j=f,Z,W,h_1,h_2}\int_{4m_j^2}^{\infty}ds\sigma_{jj\rightarrow\bar{\psi}\psi}(s)(s-4m_j^2)\sqrt{s}K_1(\dfrac{\sqrt{s}}{T}),
\end{align} 
and for the scalar DM, it reads
\begin{align} \nonumber
\label{eq:11}
\dfrac{dn_{\chi}}{dt}+3Hn_{\chi}&= \dfrac{T}{\pi^2}\sum_{i=1}^{2}m_{h_i}^2K_1(\dfrac{m_{h_i}}{T})\Gamma_{h_i\rightarrow\chi\chi} \\
&+\dfrac{T}{32\pi^4}\sum_{j=f,Z,W,h_1,h_2}\int_{4m_j^2}^{\infty}ds\sigma_{jj\rightarrow\chi\chi }(s)(s-4m_j^2)\sqrt{s}K_1(\dfrac{\sqrt{s}}{T}).
\end{align}
Here $H$ is the Hubble constant, $K_1$ is the modified Bessel function of order 1 and $s$ is the  center of mass energy squared. All contributions to the DM relic density  are considered in the corresponding cross sections and decay widths in the two above equations. Our analytical results for the cross sections and decay widths are presented in the Appendix. The number density of DM particles is calculated as $n_i=\dfrac{g_i}{(2\pi)^3}\int d^3pf_i$ \cite{Kolb:1990vq,Edsjo:1997bg,Bernal:2017kxu} (with $i=\psi, \chi$), where $f_i$ is the phase space density of particle $i$ with the $g_i$-internal spin degrees of freedom. As it is well-known from the freeze-in mechanism of production, the two DM candidates in the present model have negligible initial abundance (individually), thus we may set $f_i = 0$. Consequently, it can be derived from Eqs.~\eqref{eq:10} and (\ref{eq:11})  that the process of DM conversion, i.e. $\chi\chi\leftrightarrow\bar{\psi}\psi$, does not contribute to the total relic abundance and is suppressed in our  next calculations. On the other hand, each of the DM candidates, independent of the other, can be produced or annihilated in the Universe.

By solving Eqs.~\eqref{eq:10} and (\ref{eq:11}), one can obtain the number densities ($n_{\psi}$, $n_{\chi}$) scaled to the entropy of Universe $\hat{s}$, i.e. $Y_\psi=n_{\psi}/\hat{s}$ and $Y_\chi=n_{\chi}/\hat{s}$, as    
\begin{align} \nonumber \label{eq:14}
Y_{\psi}&= \dfrac{1}{4\pi^4}\dfrac{45M_{pl}}{1.66g^*_s(T)\sqrt{g^*_\rho}}[2\sum_{i=1}^{2}\Gamma_{h_i\rightarrow\bar{\psi}\psi}m_{h_i}^2\int_{T_{Now}}^{\infty}dT\dfrac{K_1(\dfrac{m_{h_i}}{T})}{T^5} \\
&+\sum_{j=f,Z,W,h_1,h_2}\dfrac{1}{16\pi^2}\int_{T_{Now}}^{\infty}dT\dfrac{1}{T^5}\int_{4m_j^2}^{\infty}ds\sigma_{jj\rightarrow\bar{\psi}\psi}(s)(s-4m_j^2)\sqrt{s}K_1(\dfrac{\sqrt{s}}{T})],
\end{align}
and
\begin{align} \nonumber \label{eq:15}
Y_{\chi}&= \dfrac{1}{4\pi^4}\dfrac{45M_{pl}}{1.66g^*_s(T)\sqrt{g^*_\rho}}[2\sum_{i=1}^{2}\Gamma_{h_i\rightarrow\chi\chi}m_{h_i}^2\int_{T_{Now}}^{\infty}dT\dfrac{K_1(\dfrac{m_{h_i}}{T})}{T^5} \\
&+\sum_{j=f,Z,W,h_1,h_2}\dfrac{1}{16\pi^2}\int_{T_{Now}}^{\infty}dT\dfrac{1}{T^5}\int_{4m_j^2}^{\infty}ds\sigma_{jj\rightarrow \chi\chi}(s)(s-4m_j^2)\sqrt{s}K_1(\dfrac{\sqrt{s}}{T})],
\end{align} 
where $M_{pl}$ is the Planck mass, and $g^*_s$ and $g^*_{\rho}$ are the effective numbers of degrees of freedom. 

\subsection{Relic Abundance}
\label{sec3.2}

\begin{figure}[H]
		\begin{subfigure}{0.5\textwidth}
		\includegraphics[width=1\linewidth, height=5.5cm]{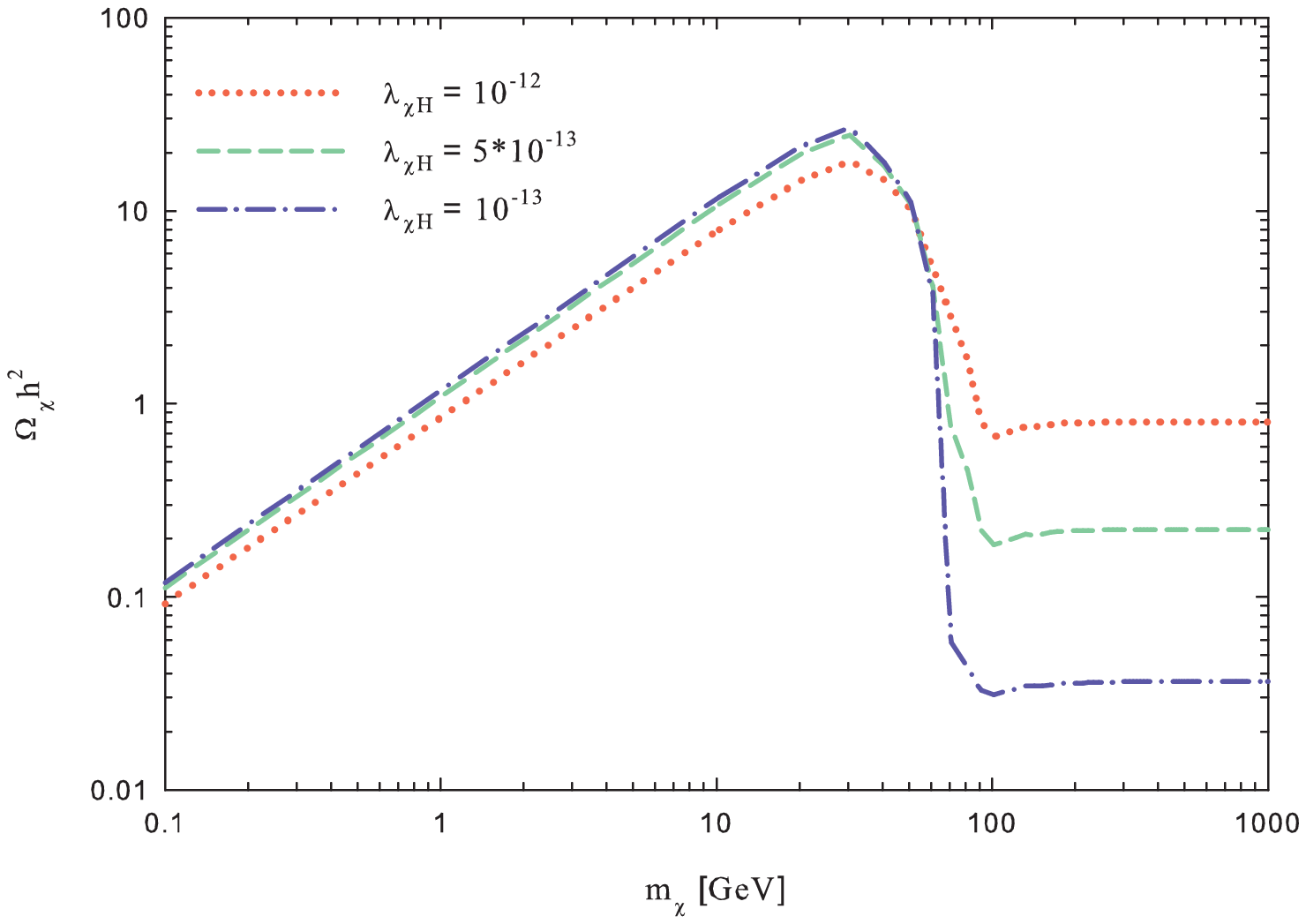} 
		\caption{}
		\label{fig:scalar1}
	\end{subfigure}
	\begin{subfigure}{0.5\textwidth}
		\includegraphics[width=1\linewidth, height=5.5cm]{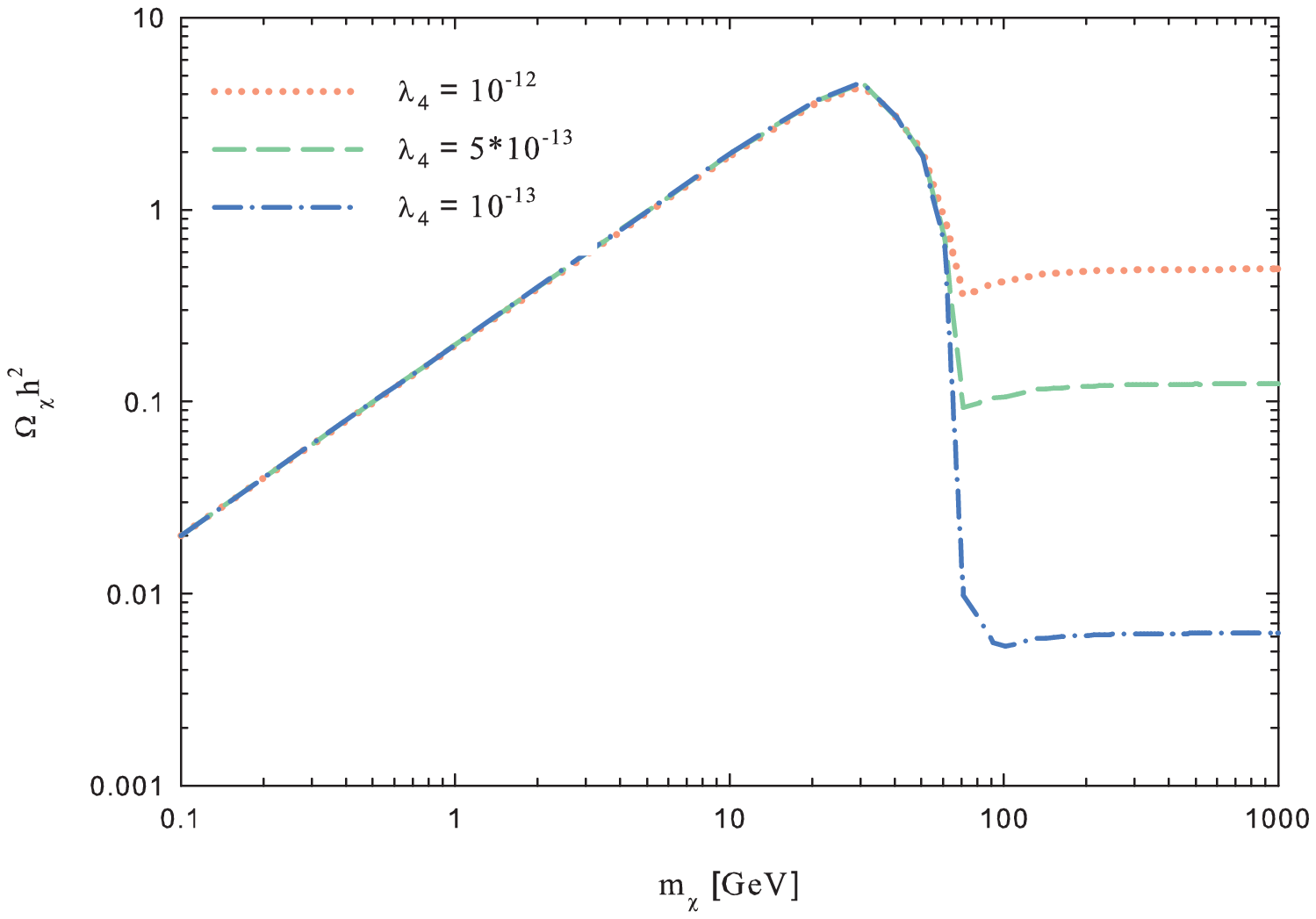}
		\caption{}
		\label{fig:scalar2}
	\end{subfigure}
	\caption{Relic density of scalar DM in terms of its mass. In this figure, we set $\lambda_3=5\times10^{-10}$, $m_{h_2}=100$~GeV and $\sin\theta=0.01$, a) for $\lambda_4=10^{-13}$ and different values of $\lambda_{\chi H}$, and b) for $\lambda_{\chi H}=10^{-13}$ and different values of $\lambda_4$.}
	\label{fig:scalar}
\end{figure}

The most important constraint which should be satisfied in models describing  DM is the observed relic  density. As the Planck experiments have measured the current amount of DM \cite{Ade:2015dga}, our first experimental constraint is described as
\begin{equation}
\Omega_\text{DM}h^2=\Omega_{\psi}h^2+\Omega_{\chi}h^2=0.1199\pm0.0027,
\end{equation}   
where $h$ is the Hubble parameter scaled in units of 100 km/s.Mpc. Using the  yield calculated in the previous section (Eqs.~\eqref{eq:14} and (\ref{eq:15})), we can obtain the relic density as  
\begin{equation}\label{eq:17}
\Omega_ih^2=2.742\times10^{-8}(\dfrac{m_i}{\text{GeV}})Y_i(T_0),~~   i=\psi, \chi.
\end{equation} 
First, we start with the scalar component of the model. The dependence of DM density is evaluated over the relevant parameters.  The predicted relic density of our model is best behaved at $m_{h_2}=100$~GeV and $\sin\theta=0.01$, with the required value of $5\times10^{-10}$~GeV for  mediator-scalar DM coupling $\lambda_3$. Two other couplings, $\lambda_{\chi H}$ and $\lambda_4$, are found to have major roles in controlling the relic density. By varying the singlet scalar DM mass (inspired by Eq.~\eqref{eq:17}), we probe our parameter space in two classes: different values of $\lambda_{\chi H}$ and of $\lambda_4$ (see Fig.~\ref{fig:scalar}).       
%%%%%%%%%%%%%%%%%%%%%%%%%%%%%%%%%%%%%%%%%%%%%%
%%%%%%%%%%%%%%%%%%%%%%%%%%%%%%%%%%%%%%%%%%%%%%
The behavior of the relic density $\Omega_\chi h^2$ regarding different Higgs-scalar DM couplings, $\lambda_{\chi H}$, is depicted  in the logarithmic scale in Fig.~\ref{fig:scalar1}. In addition to the fixed values of relevant parameters $\lambda_3$, $m_{h_2}$ and $\sin\theta$, we have adopted $\lambda_{\chi H}$ as $10^{-12}$, $5\times10^{-13}$ and $10^{-13}$  as we  fixed $\lambda_4=10^{-13}$.   Through Fig.~\ref{fig:scalar1}, it is obvious that the resonance occurs at $m_\chi\sim m_{h_2}/2$. For masses below the resonance, the relic density of the scalar component increases linearly  in the log-scale, but for larger values ($m_\chi> m_{h_2}/2$) it  seems that the relic density is independent of the mass $m_\chi$. The difference between these two regions is  due to the process $h_2\rightarrow\chi\chi$ which is allowed in the region below the resonance.         

\begin{figure}
		\begin{subfigure}{0.5\textwidth}
		\includegraphics[width=1\linewidth, height=5.5cm]{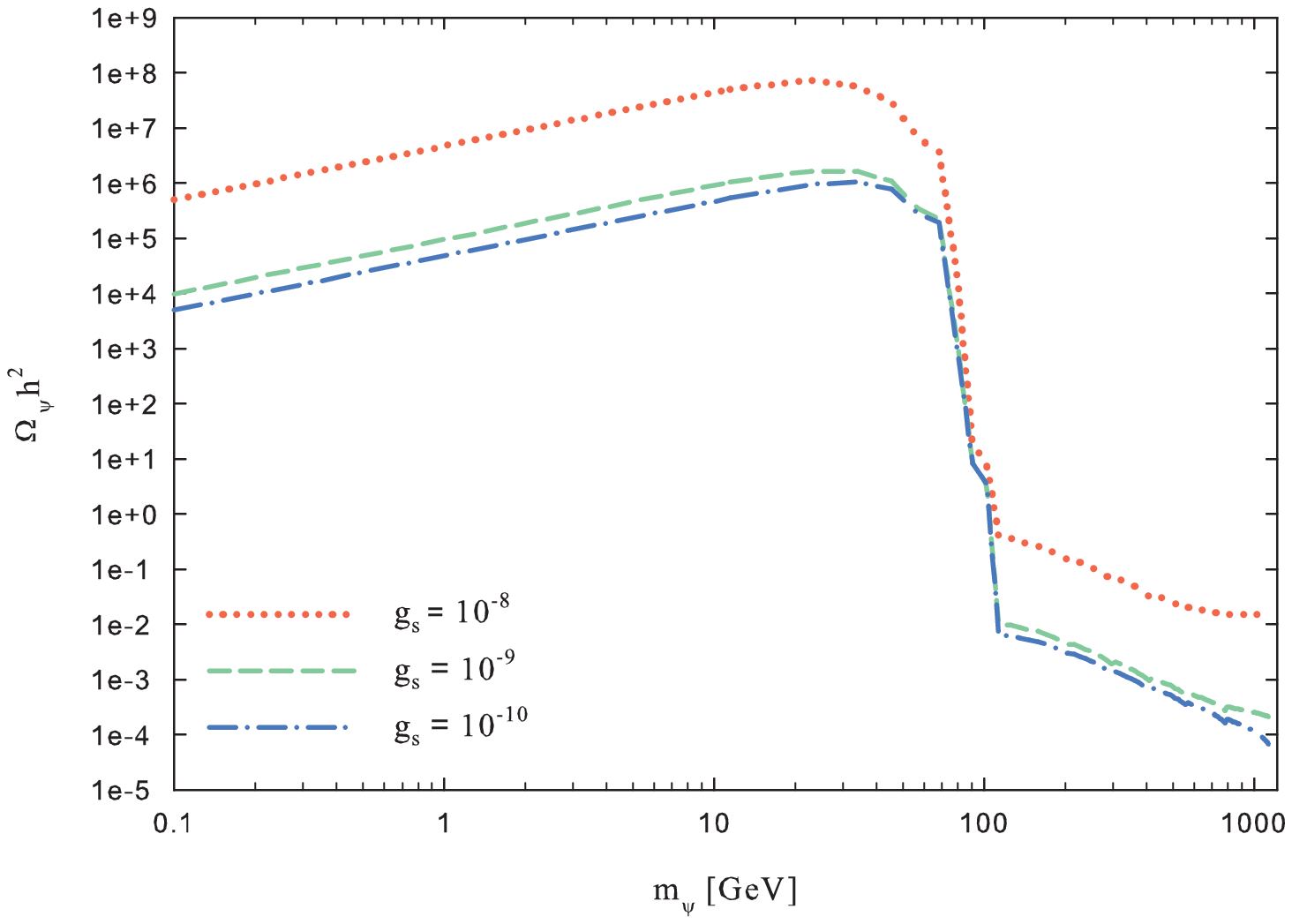} 
		\caption{}
		\label{fig:fermion1}
	\end{subfigure}
	\begin{subfigure}{0.5\textwidth}
		\includegraphics[width=1\linewidth, height=5.5cm]{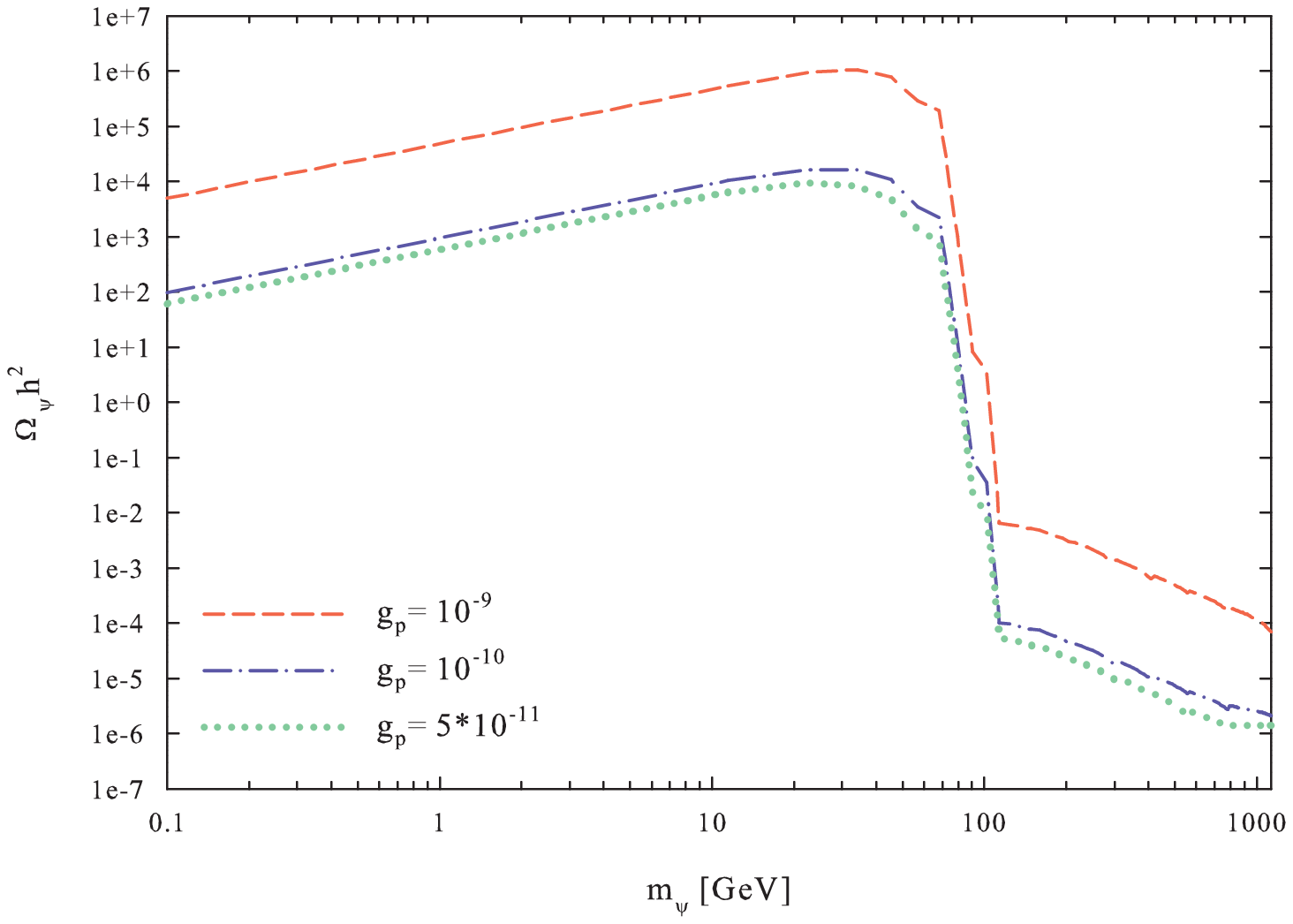}
		\caption{}
		\label{fig:fermion2}
	\end{subfigure}
	\caption{Relic density of the fermionic DM in terms of its mass. Other relevant parameters are taken as $m_{h_2}=100$~GeV and $\sin\theta=0.01$, a) for $g_p=10^{-9}$ and different values of $g_s$, and b) for $g_s=10^{-10}$ and different values of $g_p$.}
	\label{fig:fermion}
\end{figure}

The complementary analysis of the scalar component is plotted in Fig.~\ref{fig:scalar2}, where we have chosen $\lambda_4=10^{-12}, 5\times10^{-13}$ and $10^{-13}$. Regarding the resonance  at $m_\chi\sim m_{h_2}/2$, as in Fig.~\ref{fig:scalar1}, it can be seen that for the region below the resonance, the relic density grows when the scalar mass $m_\chi$ increases. This part of the graph seems to be independent of the $\lambda_4$-value and it is enhanced by the $h_2\rightarrow\chi\chi$ process. After a significant drop at $m_\chi\sim m_{h_2}/2$, the relic density seems to be  independent of DM mass for the region above the resonance. It is mainly influenced by changing the quartic coupling $\lambda_4$.  Finishing our investigation of the scalar component, it should be noted that this analysis has a good overlap with a singlet scalar model \cite{Yaguna:2011qn}.        
We continue our investigation in parallel by turning our attention to the fermionic DM  in the logarithmic scale, too. As before, we consider two classes of variations defined by the effect of scalar (parameterized by $g_s$) and pseudoscalar (parameterized by $g_p$) interactions of $\psi$ (see Fig.~\ref{fig:fermion}). We first look at the coupling $g_s$ so  its best effects are formed for the values of $10^{-8}$, $10^{-9}$ and $10^{-10}$ (Fig.~\ref{fig:fermion1}). Similar to the scalar case, the resonance position occurs at $m_\psi\sim m_{h_2}/2$, so below this value one can observe the linear behavior of relic density which arises through the process $h_2\rightarrow\psi\psi$. For $m_\psi>m_{h_2}/2$, the relic density changes by several order of magnitudes in a small interval of mass range. In this region, $\Omega_{\psi}h^2$ decreases when $m_\psi$ increases. Note that the relic density is approximately independent of DM mass for  large  values of $g_s$. Also,  for small enough values of $g_s$, decreasing $g_s$ does not significantly change the relic density.
%%%%%%%%%%%%%%%%%%%%%%%%%%%%%%%%%%%%%%%%%%%%%   
%%%%%%%%%%%%%%%%%%%%%%%%%%%%%%%%%%%%%%%%
We keep on probing our model parameter space by choosing  the appropriate values of $g_p$ as $10^{-9}$, $10^{-10}$ and $5\times10^{-11}$.  In Fig.~\ref{fig:fermion2}, we show the behavior of the relic density of the fermionic DM in terms of its mass. Here, there is a distinct point  which should be expressed. As is seen, for very small values of $g_p$, the relic density is approximately independent of mass for massive DM.  Similar to the scalar component, we can compare the fermionic component with models describing singlet fermionic DM like in Ref.~\cite{Klasen:2013ypa}.
   
\section{Phenomenological Implications}
\label{sec4}

\subsection{Direct Searches}
\label{sec4.1}

In this section, we search for signals inspired from XENON100 \cite{Aprile:2012nq} and LUX \cite{Akerib:2013tjd} in spin-independent elastic scattering of DM off nuclei. Our intended process includes the fundamental interaction of DM-quark which occurs via the t-channel mediated by scalars $h_1$ and $h_2$. Taking into account the contribution of each DM component and using fractions $\xi_\psi=\frac{\Omega_\psi}{\Omega_\text{DM}}$ and $\xi_\chi=\frac{\Omega_\chi}{\Omega_\text{DM}}$, we investigate whether the model parameter space could be affected by the experimental results in this way.  To this end, we calculate the following cross sections,  

\begin{equation}
\sigma_{SI}^\psi = \xi_\psi\frac{ g_s^2\mu_m^2\sin^2\theta\cos^2\theta}{\pi}(\dfrac{1}{m_{h_1}^2}-\dfrac{1}{m_{h_2}^2})^2\lambda_N^2,
\end{equation}
and
\begin{align} \nonumber \label{eq:19}
\sigma_{SI}^\chi &= \xi_\chi\frac{\mu_m^2}{4\pi m_{h_1}^4m_{h_2}^4m_{\chi}^2}\Big[2 v_H \lambda_{\chi H}(m_{h_1}^2\sin^2\theta+m_{h_2}^2\cos^2\theta),\\
&+\lambda_3\sin2\theta(m_{h_1}^2-m_{h_2}^2)\Big]^2\lambda_N^2,
\end{align}
where
\begin{equation}
\lambda_N = \frac{m_N}{v_H}[\sum_{q=u,d,s}f_q+\dfrac{2}{9}(1-\sum_{q=u,d,s}f_q)]\approx1.4\times10^{-3}.
\end{equation}
Here, the parameter $\mu_m=\dfrac{m_Nm_\text{DM}}{m_N+m_\text{DM}}$ is the reduced mass of DM-nucleon and $m_{\chi}=(m_{0\chi}^2+\lambda_{\chi H}v_H^2)^{1/2}$ is the physical mass of the scalar DM. A cancellation effect \cite{Esch:2014jpa} could occur when the two terms in Eq.~\eqref{eq:19} cancel each other out, giving a suppressed cross section which is not appropriate for our consideration. Generally, as was  mentioned earlier, a necessary condition for our DM candidates to be nonthermal is  that they have extremely small couplings ($g_s$, $\lambda_{\chi H}$) which would yield cross sections out of the sensitivity of the aforementioned experiments by their established values of order $\sim10^{-8}-10^{-12}$. Searching for other viable experiments, we consider the scattering of DM off free electrons in materials such as superconductors, semiconductors and graphene.  From Ref.~\cite{Hochberg:2015pha,Hoch1,Hoch2} it is seen that, although, these  electron detectors are useful for light DM particles ($\cal O$(MeV)),  the mediator mass should also be  of order $\cal O$(MeV), which is in conflict with the current model including $m_{h_2}=100$~GeV.  For this reason, we are not able to probe such FIMP models directly. This outcome is consistent with the lack of direct experimental signals to date.   

\subsection{Indirect Searches} \label{sec4.2}
\subsubsection{Invisible Higgs Decays}
\label{sec4.2.1}

%%%%%%%%%%%%%%%%%%%%%%%%%%%%%%%%%%%%%%%%%%%%%%%%
\begin{figure}
	
	\begin{subfigure}{0.5\textwidth}
		\includegraphics[width=1\linewidth, height=5.5cm]{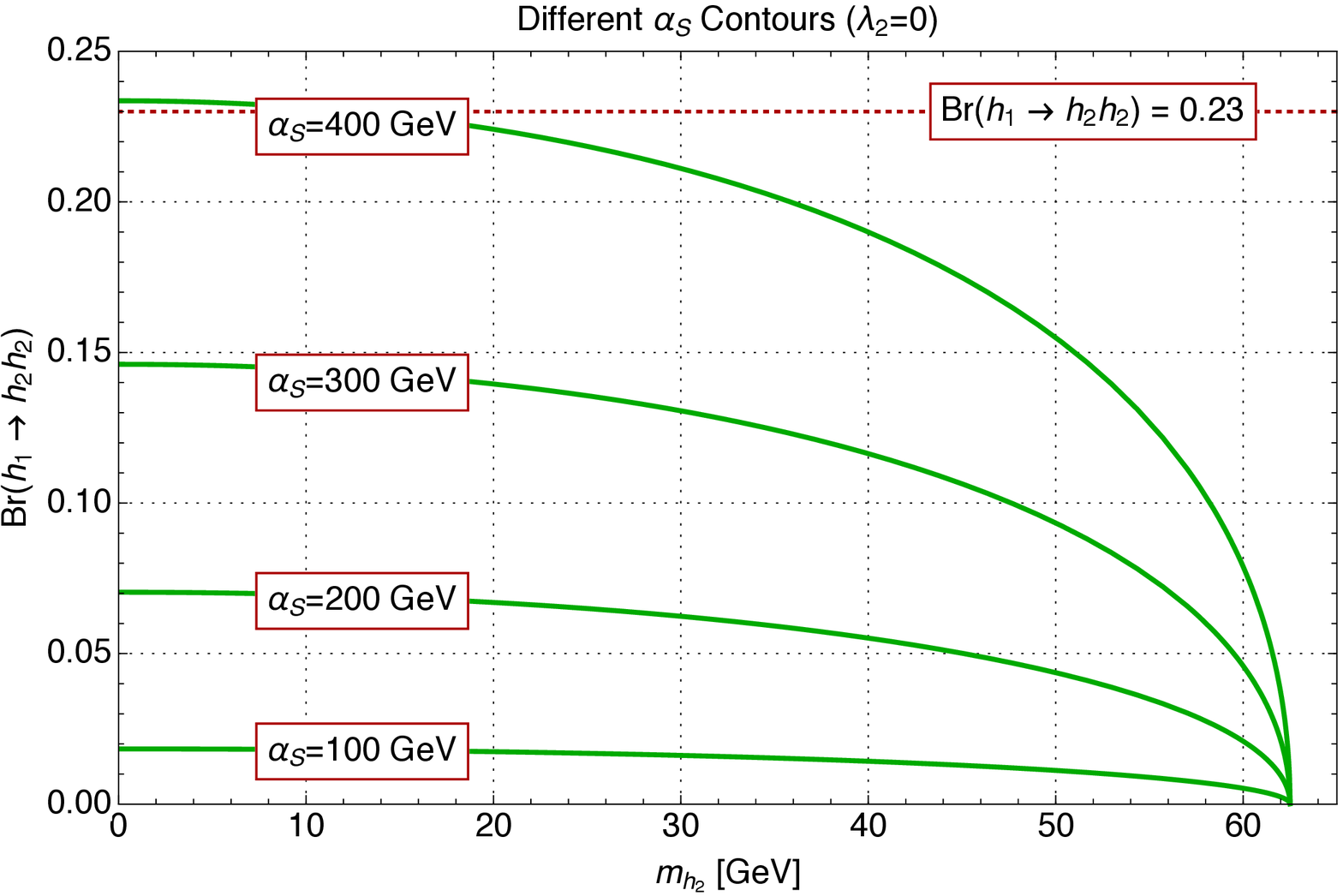} 
		\caption{}
		\label{brplotpert}
	\end{subfigure}
	\begin{subfigure}{0.5\textwidth}
		\includegraphics[width=1\linewidth, height=5.5cm]{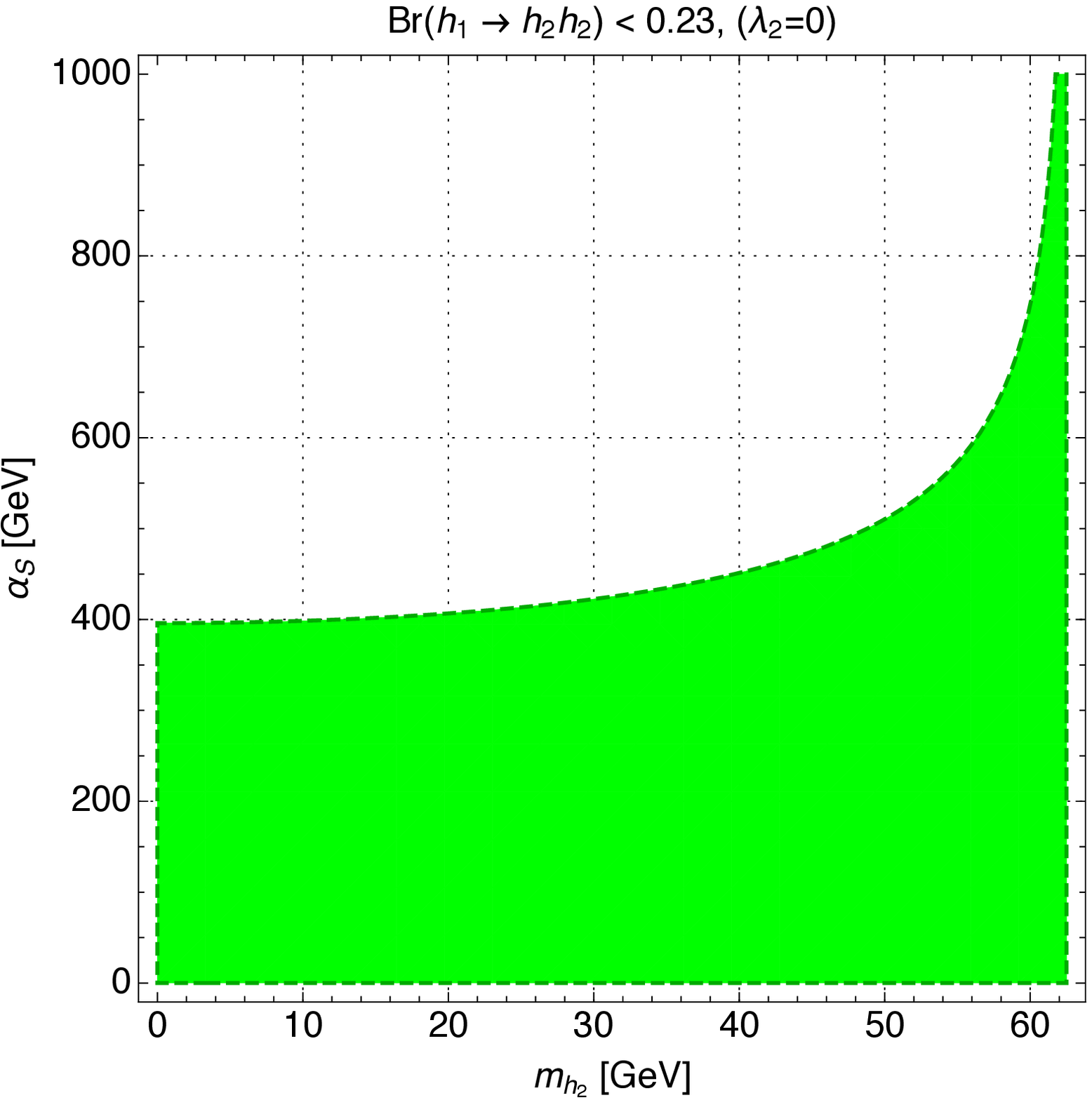}
		\caption{}
		\label{allowedBrPlot}
	\end{subfigure}
	\caption{a) Branching ratio of the SM Higgs to $h_2h_2$ for different values of $\alpha_s$ and $m_{h_2}$. The red dashed line  (at $0.23$) is the experimental bound. b) The allowed region in the $(\alpha_s, m_{h_2})$-parameter space where Br$(h_1\rightarrow h_2h_2) < 0.23$. All
		points in the green area satisfy the perturbativity condition ($|e_3| < 4\pi$).}
	\label{fig:inv}
\end{figure}
%%%%%%%%%%%%%%%%%%%%%%%%%%%%%%%%%%%%%%%%%%%%%%%%%%%%

Since the ATLAS and CMS have recorded the signature of the SM Higgs \cite{Aad:2012tfa,Chatr}, new searches have been prepared for DM phenomenology. This is done by considering the branching ratios of the Higgs, especially for decaying into light DM candidates,
\begin{equation}
\text{Br}(h_1\rightarrow\text{Invisible})=\dfrac{\Gamma^\text{inv}(h_1\rightarrow\chi\chi)+\Gamma^\text{inv}(h_1\rightarrow\psi\psi)}{\Gamma^\text{SM}+\Gamma^\text{inv}}.
\end{equation}    
Regarding the experimental upper bound $0.23$ for $\text{Br}(h_1\rightarrow\text{Invisible})$ \cite{Belanger:2013kya}, we see that the decays of Higgs to both DM $\psi$ and $\chi$ are suppressed due to small couplings $g_s$ and $g_p$ in the former case, and small $\lambda_4$ and $\lambda_{\chi H}$ in the latter. However, another constraint comes from the decay of  our Higgs to $h_2$ (if kinematically possible, i.e. $m_{h_2} < m_{h_1}/2$) whose decay rate could be calculated as 
\begin{equation}\label{gama}
\Gamma(h_1\rightarrow h_2h_2)=\dfrac{e_3^2}{8\pi m_{h_1}}(1-\dfrac{4m_{h_2}^2}{m_{h_1}^2})^{1/2} \Theta(m_{h_1}-2m_{h_2}),
\end{equation}
where $e_3$ is the relevant vertex factor which is presented in the Appendix, see Eq.~\eqref{e3}.  From Eq.~\eqref{gama}, it can be seen that for $\sin\theta=0.01$ and $m_{h_1}=125$~GeV,
the result is sensitive to the choice of $\alpha_s$.  Consequently, we investigate the behavior of the aforementioned decay rate regarding the mass of the Higgs $h_1$ and the relevant coupling $\alpha_s$.  The parameter space of our Higgs sector is plotted in Figs.~\ref{brplotpert} and \ref{allowedBrPlot} where, respectively, we have calculated $\text{Br}(h_1\rightarrow h_2h_2)$ as a function of $m_{h_2}$ and depicted the parameter space for the ($\alpha_s$, $m_{h_2}$)-plane which is consistent with experimental measurements.

\subsubsection{DM Self-Interactions}
\label{sec4.2.2}

Of the different DM models, the collision-less cold DM (CDM) paradigm has been successful in explaining the large scale structure of the Universe. However, there
are discrepancies between the CDM predictions and observations on smaller scales. The self-interacting DM (SIDM) paradigm has the potential to solve these issues (for a review of SIDM, see Ref.~\cite{Tulin:2017ara}). Although such interactions cannot be detected in experiments, we can infer bounds on $\sigma_{\text{DM}}/m$ by evaluating the trajectory of DM in colliding
galaxy clusters \cite{Clowe:2003tk, Randall:2007ph}. An updated work \cite{Kaplinghat:2015aga} has considered a set of twelve galaxies and six clusters in order to cover different scales. Including the core sizes from dwarf to cluster (varying from 0.5 to 50 kpc), the aforementioned cross section is parametrized as
\begin{equation}
\sigma_{\text{DM}}/m\sim0.1-2~ \text{cm}^2\text{g}^{-1}.
\end{equation}
In this section, we analyze this constraint to see if it can put new limits on the parameter space of our model. 

\begin{figure}[H]
	
\centering
\includegraphics[width=0.6\linewidth, height=6.3cm]{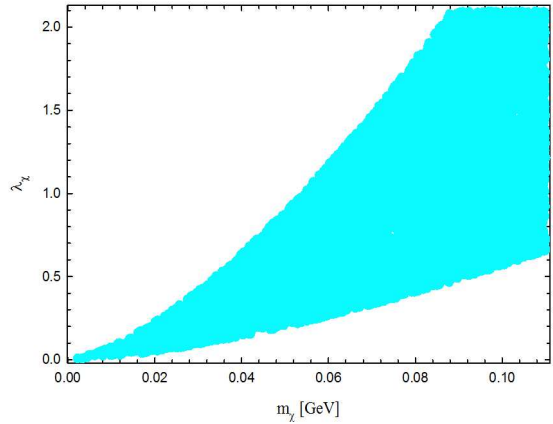}
\caption{The colored area depicts the ranges of parameter space in the ($m_\chi$, $\lambda_\chi$)-plane for the scalar DM self-interaction cross section in the range 0.1-2 cm$^2$g$^{-1}$.}
\label{selfint}
\end{figure}

The DM self-interaction in the present model includes the processes $\chi\chi\rightarrow\chi\chi$, $\psi\psi\rightarrow\psi\psi$, $\chi\chi\rightarrow\psi\psi$, $\psi\psi\rightarrow\chi\chi$ and $\chi\psi\rightarrow\chi\psi$. Except for $\chi\chi\rightarrow\chi\chi$ and $\psi\psi\rightarrow\psi\psi$, the processes contain cross sections proportional to the coupling $\lambda_4$, which is very small in our work. Therefore, the specified processes do not contribute to this cosmological constraint. Concerning the processes $\chi\chi\rightarrow\chi\chi$ and $\psi\psi\rightarrow\psi\psi$, we  start first with the scalar component which has been studied in a singlet FIMP scalar model in Ref.~\cite{Campbell:2015fra}.  Here, we just consider the contact interaction which is parameterized by the coupling $\lambda_{\chi}$. Practically, we neglect the contributions from the s-channel mediated diagrams. This is due to the small couplings of the scalar DM with both the SM Higgs and the mediator and also due to the large masses which appear in the propagator. One way to vitalize the s-channel contribution might be through fine tuning by considering the scattering near resonance (similar to Ref.~\cite{Campbell:2015fra}). In this way, in the denominator of the propagator, $m_{h_2}$ should be tuned such that $|m_\chi-m_{h_2}|\ll1$ GeV. Considering the values of couplings needed  for the observed relic density,  this scenario  fails too. Therefore, following Ref.~\cite{Campbell:2015fra}, we obtain the  self-interaction cross section per mass $m_\chi$, as
\begin{equation}
\dfrac{\sigma_\chi}{m_\chi}=\dfrac{9\lambda_{\chi}^2}{2\pi m_\chi^3},
\end{equation}
where $\lambda_{\chi}$ is the quartic self-coupling of the scalar DM (see Eq.~\eqref{eq:2}). Following the theoretical constraints in Eqs.~\eqref{eq:7} and \eqref{eq:8}, we obtain an experimental upper bound of about $0.1$~GeV on the mass of the scalar DM, which is depicted in Fig.~\ref{selfint}. Going back to Figs.~\ref{fig:scalar1} and \ref{fig:scalar2}, we observe that this range of scalar mass can produce proper total relic density along with the contribution of the fermionic component.
%%%%%%%%%%%%%%%%%%%%%%%%%%%%%
%%%%%%%%%%%%%%%%%%%%%%%%%%%%%%%%%%%%%%%%%%%%%
Another significant point, which we would like to clarify in this work, is the self-interaction of a singlet fermionic FIMP DM. In general, we have two concerns. First, it should be noted that significant self-scattering at dwarf scales requires the mediator masses to be smaller than $100$~MeV \cite{Tulin:2013teo}. In fact, following Ref.~\cite{Tulin:2013teo}, the fermionic DM should satisfy the relation  $(m_\psi /10~{\rm GeV}) (m_{h_2} / 100~\rm MeV)^2\sim1$, with Yukawa interactions of strengths $10^{-5}$ to 1, which is in contradiction with our fermionic DM coupling and mediator mass. 
Second, if the mediator couples to the SM through a Higgs portal, one should make sure that the mediator decays before the start of Big Bang nucleosynthesis (BBN), so the decay products do not affect the BBN. Eventually, we require a mediator with a lifetime $\sim 1$~s. One way to alleviate the second constraint in DM models with extremely weak interactions is to open a new decay channel for the mediator so it can decay faster. This is done
in Refs.~\cite{Kainulainen:2015sva} and \cite{Kouvaris:2014uoa} by coupling the mediator to a light sterile neutrino (it should be noted that this new coupling does not affect the relic density). However, the first constraint (light mediator) is in conflict with our mediator of mass $100$~GeV (and other usual two-component models). A promising solution seems to be to work at the resonance region  to minimize this mass constraint \cite{Duch:2017nbe}. Due to the small coupling $g_p$ (pseudoscalar interaction type), and the fact that there is an energy (velocity) dependent correction to the width in the resonance region (as explained in Ref.~\cite{Duch:2017nbe}), we conclude that the resonance DM self-interaction scenario does not work in fermionic FIMP models.               

\section{Conclusions and outlook}
\label{sec5}

We have constructed a minimal two-component model to analyze the implications of multi-component DM in the Universe. Using the freeze-in mechanism, we calculated the relic abundance predicted by our model and compared it to the observed relic abundance of DM. We started our investigation by proposing two DM particles: a real scalar and a Dirac
fermion. Furthermore, a scalar mediator between the dark sector and the SM sector was added. The couplings for this interaction are assumed to be small as we are utilizing the freeze-in
mechanism. We solved two independent Boltzmann equations in order to obtain the observed relic density with  the contributions of both DM components.
It should be noted here that at the time of finishing this work, a new version of micrOMEGAs \cite{Belanger:2018ccd} was presented  which can compute the relic abundance of FIMP candidates. In the following, using theoretical constraints, we probed the model parameter space and compared our results with the relevant singlet models. Although it is difficult to probe FIMP particles, we looked for astrophysical probes, first considering direct detection.  We considered the scattering of DM particles off nuclei and free electrons. As we explained, it is impossible to see this direct signature for our FIMP model. 

In order to constrain the parameter space of our model, we also checked the limits from the invisible decay width of the Higgs.  Finally, we probed the self-interaction of DM
in this model. We used the bounds on non-gravitational interactions of DM in giant cluster collisions and constrained the mass of DM candidates in our model. In addition to the mentioned probes of DM, we can refer to the Big Bang nucleosynthesis  and cosmic microwave background  constraints regarding dark photons and dark Higgs \cite{Fradette:2014sza, Berger:2016vxi}. These neutral bosons  mix with the SM photon (kinetically) and the SM Higgs, respectively, by accepting significant bounds on their coupling parameters. 

Two-component DM is a starting point for considering multi-component configurations where DM consists of various types of fundamental particles (scalar, fermion, vector and etc). The freeze-in framework is also a well-motivated approach which may be probed more extensively by future experiments.

\section{Acknowledgment}

We are particularly grateful to Yonit Hochberg for giving us insights into the direct probes, and we would like to thank Takashi Toma, Ian Shoemaker, Madhurima Pandey and Anirban Biswas for useful discussions.  

%%%%%%%%%%%%%%%%%%%%%%%%%%%%%%%%%%%%%%%%%%%%%%%%%%%%%%%%%%
%
\appendix

\section*{Appendix:  DM production cross sections and decay rates}

Here, we present our calculation of the fermionic DM  production cross-sections which contribute to the relic density of our  model:
%%%%%%%%%%%%%%%%%%%%%%%%%%%
\begin{eqnarray}
(\sigma v_{rel})_{ \overline{f}f\rightarrow\overline{\psi}\psi}=\frac{s\sin^2{2}\theta}{32\pi}N_c(\frac{m_f}{v_H})^2(1-\frac{4m_f^2}{s}) F
\end{eqnarray}
%%%%%%%%%%%%%%%%%%%%%%%%%
\begin{eqnarray}
(\sigma v_{rel})_{ZZ\rightarrow\overline{\psi}\psi}=\frac{\sin^2{2}\theta}{36\pi}(\frac{m_Z^2}{v_H})^2(2+\frac{(s-2m_Z^2)^2}{4m_Z^4})F
\end{eqnarray}
%%%%%%%%%%%%%%%%%%%%%%%%%%%%%%%
\begin{eqnarray}
(\sigma v_{rel})_{W^+W^-\rightarrow\overline{\psi}\psi}=\frac{\sin^2{2}\theta}{36\pi}(\frac{m_W^2}{v_H})^2(2+\frac{(s-2m_W^2)^2}{4m_W^4})F
\end{eqnarray}
where
%%%%%%%%%%%%%%%%%%%%%%%%%%%%%%%
\begin{eqnarray}
F&=&(1-\frac{4m_\psi^2}{s})^{3/2}(g_{\psi}^2+\frac{sg_p^2}{s-4m_\psi^2})\bigg\{\frac{1}{(s-m_{h_1}^2)^2+m_{h_1}^2\Gamma_{h_1}^2}+\\
&&\frac{1}{(s-m_{h_2}^2)^2+m_{h_2}^2\Gamma_{h_2}^2}-\frac{2(s-m_{h_1}^2)(s-m_{h_2}^2)+2m_{h_1}m_{h_2}\Gamma_{h_1}\Gamma_{h_2}}{[(s-m_{h_1}^2)^2+m_{h_1}^2\Gamma_{h_1}^2][(s-m_{h_2}^2)^2+m_{h_2}^2\Gamma_{h_2}^2]}\bigg\}.\nonumber
\end{eqnarray}
%%%%%%%%%%%%%%%%%%%%%%%%%%%%%
\begin{eqnarray}
(\sigma v_{rel})_{ h_1h_1\rightarrow\overline{\psi}\psi}&=&\frac{1}{4\pi s}(1-\frac{4m_\psi^2}{s})^{1/2}\Bigg\{(g_p^2s+g_\psi^2(s-4m_\psi^2))\\
&&\times \bigg[\frac{e_1^2\sin^2\theta}{(s-m_{h_1}^2)^2+m_{h_1}^2\Gamma_{h_1}^2}+\frac{e_2^2\cos^2\theta}{(s-m_{h_2}^2)^2+m_{h_2}^2\Gamma_{h_2}^2}\nonumber\\
&&+\frac{e_1e_2\sin2\theta((s-m_{h_1}^2)(s-m_{h_2}^2)+m_{h_1}m_{h_2}\Gamma_{h_1}\Gamma_{h_2})}{((s-m_{h_1}^2)^2+m_{h_1}^2\Gamma_{h_1}^2)((s-m_{h_2}^2)^2+m_{h_2}^2\Gamma_{h_2}^2)}\bigg]\nonumber\\
&&-8m_\psi g_\psi\sin^2\theta\bigg[\frac{e_1\sin\theta(s-m_{h_1}^2)}{(s-m_{h_1}^2)^2+m_{h_1}^2\Gamma_{h_1}^2}+\frac{e_2\cos\theta(s-m_{h_2}^2)}{(s-m_{h_2}^2)^2+m_{h_2}^2\Gamma_{h_2}^2}\bigg]\nonumber\\
&&\times\bigg[\frac{arctanh(x_1)[g_p^2(2m_{h_1}^2-3s)-g_\psi^2(2m_{h_1}^2-8m_\psi^2+s)]}{x_1(s-2m_{h_1})}+g_p^2-g_\psi^2\bigg]\nonumber\\
&&-4\sin^4\theta\bigg[g_p^4+g_\psi^4-2g_p^2g_\psi^2+\frac{(m_{h_1}^2(g_p^2-g_\psi^2)+4g_\psi^2m_\psi^2)^2-8g_p^2g_\psi^2m_\psi^2s}{m_{h_1}^4+m_\psi^2(s-4m_{h_1}^2)}\nonumber\\
&&-\frac{arctanh(x_1)}{x_1(s-2m_{h_1})^2}\Big(g_p^4[6m_{h_1}^4+s(s-4m_{h_1}^2)]\nonumber\\
&&-2g_p^2g_\psi^2[2m_{h_1}^2(3m_{h_1}^2-4m_\psi^2)+s(s-4m_{h_1}^2)]\nonumber\\
&&+g_\psi^4\Big[2m_{h_1}^2(3m_{h_1}^2-8m_\psi^2)+s(s-4m_{h_1}^2)+16m_\psi^2(s-2m_\psi^2)\Big]\Big)\bigg]\Bigg\},\nonumber
\end{eqnarray}
%%%%%%%%%%%%%%%%%%%%%%%%%%%%%
\begin{eqnarray}
(\sigma v_{rel})_{ h_2h_2\rightarrow\overline{\psi}\psi}&=&\frac{1}{4\pi s}(1-\frac{4m_\psi^2}{s})^{1/2}\times\Bigg\{(g_p^2s+g_\psi^2(s-4m_\psi^2))\\
&&\times \bigg[\frac{e_3^2\sin^2\theta}{(s-m_{h_1}^2)^2+m_{h_1}^2\Gamma_{h_1}^2}+\frac{e_4^2\cos^2\theta}{(s-m_{h_2}^2)^2+m_{h_2}^2\Gamma_{h_2}^2}\nonumber\\
&&+\frac{e_3e_4\sin2\theta[(s-m_{h_1}^2)(s-m_{h_2}^2)+m_{h_1}m_{h_2}\Gamma_{h_1}\Gamma_{h_2}]}{[(s-m_{h_1}^2)^2+m_{h_1}^2\Gamma_{h_1}^2][(s-m_{h_2}^2)^2+m_{h_2}^2\Gamma_{h_2}^2]}\bigg]\nonumber\\
&&-8m_\psi g_\psi\cos^2\theta\bigg[\frac{e_3\sin\theta(s-m_{h_1}^2)}{(s-m_{h_1}^2)^2+m_{h_1}^2\Gamma_{h_1}^2}+\frac{e_4\cos\theta(s-m_{h_2}^2)}{(s-m_{h_2}^2)^2+m_{h_2}^2\Gamma_{h_2}^2}\bigg]\nonumber\\
&&\times\bigg[\frac{arctanh(x_2)[g_p^2(2m_{h_2}^2-3s)-g_\psi^2(2m_{h_2}^2-8m_\psi^2+s)]}{x_2(s-2m_{h_2})}+g_p^2-g_\psi^2\bigg]\nonumber\\
&&-4\cos^4\theta\bigg[g_p^4+g_\psi^4-2g_p^2g_\psi^2+\frac{(m_{h_2}^2(g_p^2-g_\psi^2)+4g_\psi^2m_\psi^2)^2-8g_p^2g_\psi^2m_\psi^2s}{2m_{h_2}^4+2m_\psi^2(s-4m_{h_2}^2)}\nonumber\\
&&-\frac{arctanh(x_2)}{x_2(s-2m_{h_2})^2}\bigg(g_p^4(6m_{h_2}^4+s(s-4m_{h_2}^2))\nonumber\\
&&-2g_p^2g_\psi^2\bigg(2m_{h_2}^2(3m_{h_2}^2-4m_\psi^2)+s(s-4m_{h_2}^2)\bigg)\nonumber\\
&&+g_\psi^4\bigg(2m_{h_2}^2(3m_{h_2}^2-8m_\psi^2)+s(s-4m_{h_2}^2)+16m_\psi^2(s-2m_\psi^2)\bigg)\bigg)\bigg]\Bigg\},\nonumber
\end{eqnarray}
%%%%%%%%%%%%%%%%%%%%%%%%%%
\begin{eqnarray}
(\sigma v_{rel})_{ h_1h_2\rightarrow\overline{\psi}\psi}&=&\frac{1}{4\pi s}(1-\frac{4m_\psi^2}{s})^{1/2}\Bigg\{[g_p^2s+g_\psi^2(s-4m_\psi^2)]\nonumber\\
&&\times \bigg[\frac{e_2^2\sin^2\theta}{(s-m_{h_1}^2)^2+m_{h_1}^2\Gamma_{h_1}^2}+\frac{e_3^2\cos^2\theta}{(s-m_{h_2}^2)^2+m_{h_2}^2\Gamma_{h_2}^2}\nonumber\\
&&+\frac{e_2e_3\sin2\theta((s-m_{h_1}^2)(s-m_{h_2}^2)+m_{h_1}m_{h_2}\Gamma_{h_1}\Gamma_{h_2})}{[(s-m_{h_1}^2)^2+m_{h_1}^2\Gamma_{h_1}^2][(s-m_{h_2}^2)^2+m_{h_2}^2\Gamma_{h_2}^2]}\bigg]\nonumber\\
&&\hspace{-1cm}-4m_\psi g_\psi\sin2\theta\bigg[\frac{e_2\sin\theta(s-m_{h_1}^2)}{(s-m_{h_1}^2)^2+m_{h_1}^2\Gamma_{h_1}^2}+\frac{e_3\cos\theta(s-m_{h_2}^2)}{(s-m_{h_2}^2)^2+m_{h_2}^2\Gamma_{h_2}^2}\bigg]\nonumber\\
&&\hspace{-2cm}\times\bigg[\frac{arctanh(x_i)}{x_i(s-2m_{h_i}^2)}[g_p^2(m_{h_1}^2+m_{h_2}^2-3s)-g_\psi^2(m_{h_1}^2+m_{h_2}^2-8m_\psi^2+s)]+g_p^2-g_\psi^2\bigg]\nonumber\\
&&\hspace{-2cm}-\sin^22\theta\bigg[\frac{1}{2m_{h_i}^4+2m_\psi^2(s-4m_{h_i}^2)}\bigg(-g_p^4[2m_{h_i}^4+m_{h_i}^2(m_{h_j}^2-8m_\psi^2)+2m_\psi^2s]\nonumber\\
&&\hspace{-2cm}+2g_p^2g_\psi^2\Big(2m_{h_i}^4+m_{h_i}^2(m_{h_j}^2-10m_\psi^2)-2m_\psi^2(m_{h_j}^2-3s)\Big)\bigg)\nonumber\\
&&\hspace{-2cm}-g_\psi^4[(m_{h_i}^2-4m_\psi^2)(2m_{h_i}^2+m_{h_j}^2-4m_\psi^2)-2m_\psi^2s]\nonumber\\
&&\hspace{-2cm}+\frac{arctanh(x_i)}{x_i(s-2m_{h_i})^2}\bigg(g_p^4[3m_{h_i}^2+m_{h_i}^2
(4m_{h_j}^2-3s)-m_{h_j}^4-m_{h_i}^2s+s^2]\nonumber\\
&&\hspace{-2cm}+2g_p^2g_\psi^2[-3m_{h_i}^2+m_{h_i}^2(
3(4m_\psi^2+s)-4m_{h_j}^4)+m_{h_j}^4+m_{h_j}^2(s-4m_\psi^2)-s^2]\nonumber\\
&&\hspace{-2cm}+g_\psi^4[3m_{h_i}^2+m_{h_i}^2(
-3(8m_\psi^2+s)+4m_{h_j}^2)-m_{h_j}^4-m_{h_j}^2(s-8m_\psi^2)+s^2\nonumber\\
&&+16m_\psi^2(s-2m_\psi^2)]\bigg)\bigg]\Bigg\},
\end{eqnarray}
%%%%%%%%%%%%%%%%%%%%%%%%%%%%%%%%%%%%%%%%%
where the auxiliary parameters and coupling constants have the following expressions   
%%%%%%%%%%%%%%%%%%%%%%%%%%%%%%%%%%%%%%%%%
\begin{eqnarray}
x_i=\frac{(s-4m_\psi^2)^\frac{1}{2} (s-4m_{h_i}^2)^\frac{1}{2}}{(s-2m_{h_i}^2)},~~~~~~  \text{with} ~~ i, j=1,2~~ \text{and} ~i\neq j,
\end{eqnarray}
%%%%%%%%%%%%%%%%%%%%%%%%%%
\begin{equation}
e_1=-\sin^2\theta( \alpha_s\sin\theta+6v_H\lambda_2\cos\theta)
-3\cos^2\theta(\lambda_1\sin\theta+2v_H\lambda_H\cos\theta),
\end{equation}
%%%%%%%%%%%%%%%%%%%%%%%%%%
\begin{eqnarray}
e_2&=&\sin2\theta(-\frac{1}{2}\alpha_s\sin\theta+3\lambda_Hv_H\cos\theta)\nonumber\\
&&-\lambda_1\cos\theta(1-3\sin^2\theta)+2v_H\lambda_2\sin\theta(1-3\cos^2\theta),
\end{eqnarray}
%%%%%%%%%%%%%%%%%%%%%%%%%%
\begin{eqnarray}\label{e3}
e_3&=&\sin2\theta(-\frac{1}{2}\alpha_s\cos\theta-3\lambda_Hv_H\sin\theta)\nonumber\\
&&-\lambda_1\sin\theta(1-3\cos^2\theta)-2v_H\lambda_2\cos\theta(1-3\sin^2\theta),
\end{eqnarray}
%%%%%%%%%%%%%%%%%%%%%%%%%%
\begin{equation}
e_4=-\cos^2\theta( \alpha_s\cos\theta-6v_H\lambda_2\sin\theta)
-3\sin^2\theta(\lambda_1\cos\theta-2v_H\lambda_H\sin\theta).
\end{equation}
%%%%%%%%%%%%%%%%%%%%%%%%%%%%%%%%%%%%%%%%%
The scalar component will account for the DM phenomenology by the following annihilation cross sections:
%%%%%%%%%%%%%%%%%%%%%%%%%%%%%%%%%%%%%%%%%
\begin{eqnarray}
(\sigma v_{rel})_{\bar{f}f\rightarrow \chi\chi}&=&\frac{1}{16\pi}(\frac{m_f}{v_H})^2(1-\frac{4m_f^2}{s})^{3/2}N_cR
\end{eqnarray}
%%%%%%%%%%%%%%%%%%%%%%%%%%%%%%%%
\begin{eqnarray}
(\sigma v_{rel})_{ZZ\rightarrow \chi\chi}&=&\frac{1}{36\pi s}(\frac{m_Z^2}{v_H})^2(2+\frac{(s-2m_Z^2)^2}{4m_Z^4})(1-\frac{4m_Z^2}{s})^{1/2}R,
\end{eqnarray}
%%%%%%%%%%%%%%%%%%%%%%%%%%%%%%%%%%%%%%%
\begin{eqnarray}
(\sigma v_{rel})_{W^+W^-\rightarrow \chi\chi}&=&\frac{1}{36\pi s}(\frac{m_W^2}{v_H})^2(2+\frac{(s-2m_W^2)^2}{4m_W^4})(1-\frac{4m_W^2}{s})^{1/2}R\nonumber\\
\end{eqnarray}
%%%%%%%%%%%%%%%%%%%%%%%%%%%%%%%%%%%%%%%%%%%%%%%%%%
where
%%%%%%%%%%%%%%%%%%%%%%%%%%%%%%%%%%%%%%%%%
\begin{eqnarray}
R&=&\frac{e_5^2\cos^2\theta}{(s-m_{h_1}^2)^2+m_{h_1}^2\Gamma_{h_1}^2}+\frac{e_6^2\sin^2\theta}{(s-m_{h_2}^2)^2+m_{h_2}^2\Gamma_{h_2}^2}\nonumber\\
&&-\frac{e_5e_6\sin2\theta[(s-m_{h_1}^2)(s-m_{h_2}^2)+m_{h_1}m_{h_2}\Gamma_{h_1}\Gamma_{h_2}]}{[(s-m_{h_1}^2)^2+m_{h_1}^2\Gamma_{h_1}^2][(s-m_{h_2}^2)^2+m_{h_2}^2\Gamma_{h_2}^2]}.
\end{eqnarray}
%%%%%%%%%%%%%%%%%%%%%%%%%%%%%%%%%%
\begin{eqnarray}
(\sigma v_{rel})_{h_1h_1 \rightarrow \chi\chi}&=&\frac{1}{16\pi s}(1-\frac{4m_{h_1}^2}{s})^{1/2}\nonumber\\
&&\times\Bigg\{ \bigg[\frac{e_1^2e_5^2}{(s-m_{h_1}^2)^2+m_{h_1}^2\Gamma_{h_1}^2}+\frac{e_2^2e_6^2}{(s-m_{h_2}^2)^2+m_{h_2}^2\Gamma_{h_2}^2}\nonumber\\
&&+\frac{2e_1e_2e_5e_6[(s-m_{h_1}^2)(s-m_{h_2}^2)+m_{h_1}m_{h_2}\Gamma_{h_1}\Gamma_{h_2}]}{[(s-m_{h_1}^2)^2+m_{h_1}^2\Gamma_{h_1}^2][(s-m_{h_2}^2)^2+m_{h_2}^2\Gamma_{h_2}^2]}\bigg]\nonumber\\
&&-\Big(\dfrac{8e_5^2}{s-2m_{h_1}^2}F(y_1)+2e_7\Big)\bigg[\frac{e_1e_5(s-m_{h_1}^2)}{(s-m_{h_1}^2)^2+m_{h_1}^2\Gamma_{h_1}^2}\nonumber\\
&&+\frac{e_2e_6(s-m_{h_2}^2)}{(s-m_{h_2}^2)^2+m_{h_2}^2\Gamma_{h_2}^2}\bigg]\nonumber\\
&&+\dfrac{8e_5^2}{s-2m_{h_1}^2}F(y_1)\bigg[\dfrac{e_5^2}{s-2m_{h_1}^2}\Big[\dfrac{1}{F(y_1)(1-y_1^2)}+1\Big]+e_7\bigg]
+e_7^2\Bigg\},\nonumber\\
\end{eqnarray}
%%%%%%%%%%%%%%%%%%%%%%%%%%%%%%%%%%%%%
\begin{eqnarray}
(\sigma v_{rel})_{h_2h_2 \rightarrow \chi\chi}&=&\frac{1}{16\pi s}(1-\frac{4m_{h_2}^2}{s})^{1/2}\nonumber\\
&&\times\Bigg\{ \bigg[\frac{e_3^2e_5^2}{(s-m_{h_1}^2)^2+m_{h_1}^2\Gamma_{h_1}^2}+\frac{e_4^2e_6^2}{(s-m_{h_2}^2)^2+m_{h_2}^2\Gamma_{h_2}^2}\nonumber\\
&&+\frac{2e_3e_4e_5e_6[(s-m_{h_1}^2)(s-m_{h_2}^2)+m_{h_1}m_{h_2}\Gamma_{h_1}\Gamma_{h_2}]}{[(s-m_{h_1}^2)^2+m_{h_1}^2\Gamma_{h_1}^2][(s-m_{h_2}^2)^2+m_{h_2}^2\Gamma_{h_2}^2]}\bigg]\nonumber\\
&&-\bigg(\dfrac{8e_6^2}{s-2m_{h_2}^2}F(y_2)+2e_8\bigg)\bigg[\frac{e_3e_5(s-m_{h_1}^2)}{(s-m_{h_1}^2)^2+m_{h_1}^2\Gamma_{h_1}^2}\nonumber\\
&&+\frac{e_4e_6(s-m_{h_2}^2)}{(s-m_{h_2}^2)^2+m_{h_2}^2\Gamma_{h_2}^2}\bigg]\nonumber\\
&&+\dfrac{8e_6^2}{s-2m_{h_2}^2}F(y_2)\bigg[\dfrac{e_6^2}{s-2m_{h_2}^2}(\dfrac{1}{F(y_2)(1-y_2^2)}+1)+e_8\bigg]
+e_8^2\Bigg\},\nonumber\\
\end{eqnarray}
%%%%%%%%%%%%%%%%%%%%%%%%%%%%%%%%%%%%%%%%%
\begin{eqnarray}
(\sigma v_{rel})_{ h_1h_2\rightarrow \chi\chi}&=&\frac{1}{16\pi s}(1-\frac{4m_{h_i}^2}{s})^{1/2}\nonumber\\
&&\times\Bigg\{ \bigg[\frac{e_2^2e_5^2}{(s-m_{h_1}^2)^2+m_{h_1}^2\Gamma_{h_1}^2}+\frac{e_3^2e_6^2}{(s-m_{h_2}^2)^2+m_{h_2}^2\Gamma_{h_2}^2}\nonumber\\
&&+\frac{2e_2e_3e_5e_6[(s-m_{h_1}^2)(s-m_{h_2}^2)+m_{h_1}m_{h_2}\Gamma_{h_1}\Gamma_{h_2}]}{[(s-m_{h_1}^2)^2+m_{h_1}^2\Gamma_{h_1}^2][(s-m_{h_2}^2)^2+m_{h_2}^2\Gamma_{h_2}^2]}\bigg]\nonumber\\
&&-\bigg(\dfrac{8e_5e_6}{s-2m_{h_i}^2}F(y_i)+2e_9\bigg)\bigg[\frac{e_2e_5(s-m_{h_1}^2)}{(s-m_{h_1}^2)^2+m_{h_1}^2\Gamma_{h_1}^2}\nonumber\\
&&+\frac{e_3e_6(s-m_{h_2}^2)}{(s-m_{h_2}^2)^2+m_{h_2}^2\Gamma_{h_2}^2}\bigg]\nonumber\\
&&+\dfrac{8e_5e_6}{s-2m_{h_i}^2}F(y_i)\bigg[\dfrac{e_5e_6}{s-2m_{h_i}^2}(\dfrac{1}{F(y_i)(1-y_i^2)}+1)+e_9\bigg]
+e_9^2\Bigg\},\nonumber\\
\end{eqnarray}
%%%%%%%%%%%%%%%%%%%%%%%%%%%%%%%%%%%%%%%%%
where we have employed parameter $y_i$ and function $F(y_i)$ as
%%%%%%%%%%%%%%%%%%%%%%%%%%%%%%%%%%%%%%%%%
\begin{eqnarray}
&&y_i=\frac{(s-4m_{\chi}^2)^\frac{1}{2} (s-4m_{h_i}^2)^\frac{1}{2}}{(s-2m_{h_i}^2)}  \nonumber\\
&&F(y_i)=\frac{1}{y_i}arctanh(y_i) ~~~~~~  \text{with} ~~ i=1,2~,
\end{eqnarray}
%%%%%%%%%%%%%%%%%%%%%%%%%%%%%%%%%%%%%%%%%
and also for coupling constants, we have the following parameters:
%%%%%%%%%%%%%%%%%%%%%%%%%%%%%%%%%%%%%%%%%
\begin{equation}
e_5=-[v_H\lambda_{\chi H}\cos\theta+\lambda_3\sin\theta]\times2!,
\end{equation}
%%%%%%%%%%%%%%%%%%%%%%%%%%%%%%%%%%%%%%%%%
\begin{equation}
e_6=[v_H\lambda_{\chi H}\sin\theta-\lambda_3\cos\theta]\times2!,
\end{equation}
%%%%%%%%%%%%%%%%%%%%%%%%%%%%%%%%%%%%%%%%%
\begin{equation}
e_7=-[\frac{1}{2}\lambda_{\chi H}\cos^2\theta+\lambda_4\sin^2\theta]\times4!,
\end{equation}
%%%%%%%%%%%%%%%%%%%%%%%%%%%%%%%%%%%%%%%%%
\begin{equation}
e_8=-[\frac{1}{2}\lambda_{\chi H}\sin^2\theta+\lambda_4\cos^2\theta]\times4!,
\end{equation}
%%%%%%%%%%%%%%%%%%%%%%%%%%%%%%%%%%%%%%%%%
\begin{equation}
e_9=[(\frac{1}{2}\lambda_{\chi H}-\lambda_4)\sin2\theta]\times2!.
\end{equation}
%%%%%%%%%%%%%%%%%%%%%%%%%%%%%%%%%%%%%%%%%
Finally, the decay rates of scalars $h_i$ (with $i=1,2$) into fermionic and scalar DM particles are given as: 
%%%%%%%%%%%%%%%%%%%%%%%%%%%%%%%%%%%%%%%%%
\begin{eqnarray}
\Gamma(h_i\rightarrow \bar{\psi}\psi)=\frac{S_i^2\theta}{8\pi }(1-4m_{\psi}^2/m_{h_i}^2)^\frac{3}{2}[g_\psi^2+\frac{m_{h_i}^2g_p^2}{m_{h_i}^2-4m_\psi^2}],
\end{eqnarray}
%%%%%%%%%%%%%%%%%%%%%%%%%%%%%%%%%%%%%%%%%
\begin{eqnarray}
\Gamma(h_1\rightarrow \chi\chi)=\frac{e_5^2}{32\pi m_{h_1}}(1-4m_\chi^2/m_{h_1}^2)^\frac{1}{2},
\end{eqnarray}
%%%%%%%%%%%%%%%%%%%%%%%%%%%%%%%%%%%%%%%%%
\begin{eqnarray}
\Gamma(h_2\rightarrow \chi\chi)=\frac{e_6^2}{32\pi m_{h_2}}(1-4m_\chi^2/m_{h_2}^2)^\frac{1}{2},
\end{eqnarray}
where we have defined $S_1\theta=\sin\theta$ and $S_2\theta=\cos\theta$.


\begin{thebibliography}{10}
%\cite{1}
%\cite{Gondolo:1990dk}
\bibitem{Gondolo:1990dk}
P.~Gondolo, and G.~Gelmini,
%``Cosmic abundances of stable particles: Improved analysis,''
Nucl.\ Phys.\ B, {\bf 360}:145 (1991).
%doi:10.1016/0550-3213(91)90438-4
%%CITATION = doi:10.1016/0550-3213(91)90438-4;%%
%850 citations counted in INSPIRE as of 06 Jan 2018

\bibitem{Srednickii}
M.~Srednicki, R.~Watkins, and K.~A.~Olive,
%``Calculations of Relic Densities in the Early Universe,''
Nucl.\ Phys.\ B, {\bf 310}:693 (1988).
%doi:10.1016/0550-3213(88)90099-5
%%CITATION = doi:10.1016/0550-3213(88)90099-5;%%
%305 citations counted in INSPIRE as of 06 Jan 2018

\bibitem{Chiu}
H.~Y.~Chiu,
%``Symmetry between particle and anti-particle populations in the universe,''
Phys.\ Rev.\ Lett,\  {\bf 17}:712 (1966).
%doi:10.1103/PhysRevLett.17.712
%%CITATION = doi:10.1103/PhysRevLett.17.712;%%
%105 citations counted in INSPIRE as of 04 Jan 2018

%\cite{Jungman:1995df}
\bibitem{Jungman:1995df}
G.~Jungman, M.~Kamionkowski, and K.~Griest,
%``Supersymmetric dark matter,''
Phys.\ Rept,\  {\bf 267}:195 (1996).
%doi:10.1016/0370-1573(95)00058-5 [hep-ph/9506380].
%%CITATION = doi:10.1016/0370-1573(95)00058-5;%%
%3331 citations counted in INSPIRE as of 04 Jan 2018


%\cite{Cheng:2002ej}
\bibitem{Cheng:2002ej}
H.~C.~Cheng, J.~L.~Feng, and K.~T.~Matchev,
%``Kaluza-Klein dark matter,''
Phys.\ Rev.\ Lett,\  {\bf 89}:211301 (2002).
%doi:10.1103/PhysRevLett.89.211301
%[hep-ph/0207125].
%%CITATION = doi:10.1103/PhysRevLett.89.211301;%%
%567 citations counted in INSPIRE as of 04 Jan 2018
%\cite{Servant:2002aq}

\bibitem{Servant:2002aq}
G.~Servant, and T.~M.~P.~Tait,
%``Is the lightest Kaluza-Klein particle a viable dark matter candidate?,''
Nucl.\ Phys.\ B, {\bf 650}:391 (2003).
%doi:10.1016/S0550-3213(02)01012-X
%[hep-ph/0206071].
%%CITATION = doi:10.1016/S0550-3213(02)01012-X;%%
%832 citations counted in INSPIRE as of 04 Jan 2018


%\cite{Silveira:1985rk}
\bibitem{Silveira:1985rk}
V.~Silveira, and A.~Zee,
%``Scalar Phantoms,''
Phys.\ Lett. B,\  {\bf 161}:136 (1985).
%doi:10.1016/0370-2693(85)90624-0
%%CITATION = doi:10.1016/0370-2693(85)90624-0;%%
%467 citations counted in INSPIRE as of 04 Jan 2018

\bibitem{McDonald}
J.~McDonald,
%``Gauge singlet scalars as cold dark matter,''
Phys.\ Rev.\ D, {\bf 50}: 3637 (1994).
%doi:10.1103/PhysRevD.50.3637
%[hep-ph/0702143 [HEP-PH]].
%%CITATION = doi:10.1103/PhysRevD.50.3637;%%
%586 citations counted in INSPIRE as of 04 Jan 2018

\bibitem{Burgess}
C.~P.~Burgess, M.~Pospelov, and T.~ter Veldhuis,
%``The Minimal model of nonbaryonic dark matter: A Singlet scalar,''
Nucl.\ Phys.\ B, {\bf 619}:709 (2001).
%doi:10.1016/S0550-3213(01)00513-2
%[hep-ph/0011335].


\bibitem{Barger}
V.~Barger, P.~Langacker, M.~McCaskey et al,
%``LHC Phenomenology of an Extended Standard Model with a Real Scalar Singlet,''
Phys.\ Rev.\ D, {\bf 77}:035005 (2008).
%doi:10.1103/PhysRevD.77.035005
%[arXiv:0706.4311 [hep-ph]].
%%CITATION = doi:10.1103/PhysRevD.77.035005;%%
%362 citations counted in INSPIRE as of 04 Jan 2018

%\cite{Kim:2008pp}
\bibitem{Kim:2008pp}
Y.~G.~Kim, K.~Y.~Lee, and S.~Shin,
%``Singlet fermionic dark matter,''
JHEP, {\bf 0805}:100 (2008).
%doi:10.1088/1126-6708/2008/05/100
%[arXiv:0803.2932 [hep-ph]].
%%CITATION = doi:10.1088/1126-6708/2008/05/100;%%
%138 citations counted in INSPIRE as of 04 Jan 2018

\bibitem{Ettefaghi}
M.~M.~Ettefaghi, and R.~Moazzemi,
%``Annihilation of singlet fermionic dark matter into two photons,''
JCAP, {\bf 1302}:048 (2013).
%doi:10.1088/1475-7516/2013/02/048
%[arXiv:1301.4892 [hep-ph]].
%%CITATION = doi:10.1088/1475-7516/2013/02/048;%%
%11 citations counted in INSPIRE as of 04 Jan 2018



\bibitem{Fairbairn}
M.~Fairbairn, and R.~Hogan,
%``Singlet Fermionic Dark Matter and the Electroweak Phase Transition,''
JHEP, {\bf 1309}:022 (2013).
%doi:10.1007/JHEP09(2013)022
%[arXiv:1305.3452 [hep-ph]].
%%CITATION = doi:10.1007/JHEP09(2013)022;%%
%44 citations counted in INSPIRE as of 04 Jan 2018


%\cite{McDonald:2001vt}
\bibitem{McDonald:2001vt}
J.~McDonald,
%``Thermally generated gauge singlet scalars as selfinteracting dark matter,''
Phys.\ Rev.\ Lett,\  {\bf 88}:091304 (2002).
%doi:10.1103/PhysRevLett.88.091304
%[hep-ph/0106249].
%%CITATION = doi:10.1103/PhysRevLett.88.091304;%%
%165 citations counted in INSPIRE as of 04 Jan 2018



%\cite{Hall:2009bx}
\bibitem{Hall:2009bx}
L.~J.~Hall, K.~Jedamzik, J.~March-Russell, and S.~M.~West,
%``Freeze-In Production of FIMP Dark Matter,''
JHEP, {\bf 1003}:080 (2010).
%doi:10.1007/JHEP03(2010)080
%[arXiv:0911.1120 [hep-ph]].
%%CITATION = doi:10.1007/JHEP03(2010)080;%%
%272 citations counted in INSPIRE as of 04 Jan 2018

%\cite{Bernal:2017kxu}
\bibitem{Bernal:2017kxu}
N.~Bernal, M.~Heikinheimo, T.~Tenkanen, K.~Tuominen and V.~Vaskonen,
%``The Dawn of FIMP Dark Matter: A Review of Models and Constraints,''
Int.\ J.\ Mod.\ Phys.\ A {\bf 32} (2017) no.27,  1730023
doi:10.1142/S0217751X1730023X
[arXiv:1706.07442 [hep-ph]].
%%CITATION = doi:10.1142/S0217751X1730023X;%%
%32 citations counted in INSPIRE as of 11 Mar 2018


%\cite{Yaguna:2011qn}
\bibitem{Yaguna:2011qn}
 C.~E.~Yaguna,
%``The Singlet Scalar as FIMP Dark Matter,''
JHEP, {\bf 1108}:060 (2011).
%doi:10.1007/JHEP08(2011)060
%[arXiv:1105.1654 [hep-ph]].
%%CITATION = doi:10.1007/JHEP08(2011)060;%%
%46 citations counted in INSPIRE as of 04 Jan 2018


%\cite{Klasen:2013ypa}
\bibitem{Klasen:2013ypa}
 M.~Klasen, and C.~E.~Yaguna,
%``Warm and cold fermionic dark matter via freeze-in,''
JCAP, {\bf 1311}:039 (2013).
%doi:10.1088/1475-7516/2013/11/039
%[arXiv:1309.2777 [hep-ph]].
%%CITATION = doi:10.1088/1475-7516/2013/11/039;%%
%31 citations counted in INSPIRE as of 04 Jan 2018


%\cite{Ayazi:2015jij}
\bibitem{Ayazi:2015jij}
S.~Yaser Ayazi, S.~M.~Firouzabadi, and S.~P.~Zakeri,
%``Freeze-in production of Fermionic Dark Matter with Pseudo-scalar and Phenomenological Aspects,''
J.\ Phys.\ G, {\bf 43} (9):095006 (2016).
%doi:10.1088/0954-3899/43/9/095006
%[arXiv:1511.07736 [hep-ph]].
%%CITATION = doi:10.1088/0954-3899/43/9/095006;%%
%8 citations counted in INSPIRE as of 04 Jan 2018



%\cite{Merle:2014xpa}
\bibitem{Merle:2014xpa}
A.~Merle, and A.~Schneider,
%``Production of Sterile Neutrino Dark Matter and the 3.5 keV line,''
Phys.\ Lett.\ B, {\bf 749}:283 (2015).
%doi:10.1016/j.physletb.2015.07.080
%[arXiv:1409.6311 [hep-ph]].


\bibitem{Merle}
A.~Merle, and M.~Totzauer,
%``keV Sterile Neutrino Dark Matter from Singlet Scalar Decays: Basic Concepts and Subtle Features,''
JCAP, {\bf 1506}:011 (2015).
%doi:10.1088/1475-7516/2015/06/011
%[arXiv:1502.01011 [hep-ph]].

\bibitem{Shakya}
B.~Shakya,
%``Sterile Neutrino Dark Matter from Freeze-In,''
Mod.\ Phys.\ Lett.\ A, {\bf 31} (06):1630005 (2016).
%doi:10.1142/S0217732316300056
%[arXiv:1512.02751 [hep-ph]].

%\cite{Kang:2014cia}
\bibitem{Kang:2014cia}
Z.~Kang,
%``Upgrading sterile neutrino dark matter to FI$m$P using scale invariance,''
Eur.\ Phys.\ J.\ C {\bf 75} (2015) no.10,  471
doi:10.1140/epjc/s10052-015-3702-4
[arXiv:1411.2773 [hep-ph]].
%%CITATION = doi:10.1140/epjc/s10052-015-3702-4;%%
%46 citations counted in INSPIRE as of 11 Mar 2018

\bibitem{Biswas}
A.~Biswas, and A.~Gupta,
%``Freeze-in Production of Sterile Neutrino Dark Matter in U(1)$_{\rm B-L}$ Model,''
JCAP, {\bf 1609}:044 (2016).
%Addendum: [JCAP {\bf 1705} (2017) no.05,  A01].
%doi:10.1088/1475-7516/2017/05/A01, 10.1088/1475-7516/2016/09/044
%[arXiv:1607.01469 [hep-ph]].
%%CITATION = doi:10.1088/1475-7516/2017/05/A01, 10.1088/1475-7516/2016/09/044;%%
%14 citations counted in INSPIRE as of 04 Jan 2018

%\cite{Pandey:2017quk}
\bibitem{Pandey:2017quk}
M.~Pandey, D.~Majumdar, and K.~P.~Modak,
%``Two Component Feebly Interacting Massive Particle (FIMP) Dark Matter,''
arXiv: hepph/1709.05955.
%%CITATION = ARXIV:1709.05955;%%

%\cite{Dev:2013yza}
\bibitem{Dev:2013yza}
P.~S.~Bhupal Dev, A.~Mazumdar and S.~Qutub,
%``Constraining Non-thermal and Thermal properties of Dark Matter,''
Front.\ in Phys.\  {\bf 2} (2014) 26
doi:10.3389/fphy.2014.00026
[arXiv:1311.5297 [hep-ph]].
%%CITATION = doi:10.3389/fphy.2014.00026;%%
%40 citations counted in INSPIRE as of 05 Mar 2018

%\cite{Profumo:2009tb}
\bibitem{Profumo:2009tb}
S.~Profumo, K.~Sigurdson, and L.~Ubaldi,
%``Can we discover multi-component WIMP dark matter?,''
JCAP, {\bf 0912}:016 (2009).
%doi:10.1088/1475-7516/2009/12/016;
%[arXiv:0907.4374 [hep-ph]];

\bibitem{Gelm1}
G.~B.~Gelmini,
%``Dark matter searches: Looking for the cake or its frosting? Detectability of a subdominant component of the CDM,''
Nucl.\ Phys.\ Proc.\ Suppl,\  {\bf 138}:32 (2005).
%doi:10.1016/j.nuclphysbps.2004.11.006
%[hep-ph/0310022];

\bibitem{Gelm2}
G.~Duda, G.~Gelmini, P.~Gondolo et al,
%``Indirect detection of a subdominant density component of cold dark matter,''
Phys.\ Rev.\ D, {\bf 67}:023505 (2003).
%doi:10.1103/PhysRevD.67.023505
%[hep-ph/0209266];

\bibitem{Dudaa}
G.~Duda, G.~Gelmini, and P.~Gondolo,
%``Detection of a subdominant density component of cold dark matter,''
Phys.\ Lett.\ B, {\bf 529}:187 (2002).
%doi:10.1016/S0370-2693(02)01266-2
%[hep-ph/0102200].


%\cite{Herrero-Garcia:2017vrl}
\bibitem{Herrero-Garcia:2017vrl}
J.~Herrero-Garcia, A.~Scaffidi, M.~White et al,
%``On the direct detection of multi-component dark matter: sensitivity studies and parameter estimation,''
JCAP, {\bf 1711}:021  (2017).
%doi:10.1088/1475-7516/2017/11/021
%[arXiv:1709.01945 [hep-ph]].


%\cite{Biswas:2015sva}
\bibitem{Biswas:2015sva}
 A.~Biswas, D.~Majumdar, and P.~Roy,
%``Nonthermal two component dark matter model for Fermi-LAT γ-ray excess and 3.55 keV X-ray line,''
JHEP, {\bf 1504}:065 (2015).
%doi:10.1007/JHEP04(2015)065
%[arXiv:1501.02666 [hep-ph]].


%\cite{Esch:2014jpa}
\bibitem{Esch:2014jpa}
S.~Esch, M.~Klasen, and C.~E.~Yaguna,
%``A minimal model for two-component dark matter,''
JHEP, {\bf 1409}:108 (2014).
%doi:10.1007/JHEP09(2014)108
%[arXiv:1406.0617 [hep-ph]].
%%CITATION = doi:10.1007/JHEP09(2014)108;%%
%15 citations counted in INSPIRE as of 04 Jan 2018


%\cite{Bhattacharya:2013hva}
\bibitem{Bhattacharya:2013hva}
 S.~Bhattacharya, A.~Drozd, B.~Grzadkowski et al,
%``Two-Component Dark Matter,''
JHEP, {\bf 1310}:158 (2013).
%doi:10.1007/JHEP10(2013)158
%[arXiv:1309.2986 [hep-ph]].
%%CITATION = doi:10.1007/JHEP10(2013)158;%%
%31 citations counted in INSPIRE as of 04 Jan 2018


%\cite{Biswas:2013nn}
\bibitem{Biswas:2013nn}
 A.~Biswas, D.~Majumdar, A.~Sil et al,
%``Two Component Dark Matter : A Possible Explanation of 130 GeV $\gamma-$ Ray Line from the Galactic Centre,''
JCAP, {\bf 1312}:049 (2013).
%doi:10.1088/1475-7516/2013/12/049
%[arXiv:1301.3668 [hep-ph]].
%%CITATION = doi:10.1088/1475-7516/2013/12/049;%%
%23 citations counted in INSPIRE as of 04 Jan 2018

%\cite{Chialva:2012rq}
\bibitem{Chialva:2012rq}
D.~Chialva, P.~S.~B.~Dev and A.~Mazumdar,
%``Multiple dark matter scenarios from ubiquitous stringy throats,''
Phys.\ Rev.\ D {\bf 87} (2013) no.6,  063522
doi:10.1103/PhysRevD.87.063522
[arXiv:1211.0250 [hep-ph]].
%%CITATION = doi:10.1103/PhysRevD.87.063522;%%
%38 citations counted in INSPIRE as of 05 Mar 2018

%\cite{DuttaBanik:2016jzv}
\bibitem{DuttaBanik:2016jzv}
A.~Dutta Banik, M.~Pandey, D.~Majumdar et al,
%``Two component WIMP–FImP dark matter model with singlet fermion, scalar and pseudo scalar,''
Eur.\ Phys.\ J.\ C, {\bf 77} (10):657 (2017).
%doi:10.1140/epjc/s10052-017-5221-y
%[arXiv:1612.08621 [hep-ph]].



%\cite{Modak:2014vva}
\bibitem{Modak:2014vva}
K.~P.~Modak,
%``3.5 keV X-ray Line Signal from Decay of Right-Handed Neutrino due to Transition Magnetic Moment,''
JHEP, {\bf 1503}:064 (2015).
%doi:10.1007/JHEP03(2015)064
%[arXiv:1404.3676 [hep-ph]].



%\cite{Babu:2014pxa}
\bibitem{Babu:2014pxa}
K.~S.~Babu, and R.~N.~Mohapatra,
%``7 keV Scalar Dark Matter and the Anomalous Galactic X-ray Spectrum,''
Phys.\ Rev.\ D, {\bf 89}:115011 (2014).
%doi:10.1103/PhysRevD.89.115011
%[arXiv:1404.2220 [hep-ph]].


%\cite{Ade:2015dga}
\bibitem{Ade:2015dga}
 P.~A.~R.~Ade  et al (Planck Collaboration),
%``Planck 2013 results. XXXI. Consistency of the Planck data,''
Astron.\ Astrophys,\  {\bf 571}:A31 (2014).
%doi:10.1051/0004-6361/201423743
%[arXiv:1508.03375 [astro-ph.CO]].


%\cite{Kolb:1990vq}
\bibitem{Kolb:1990vq}
E.~W.~Kolb and M.~S.~Turner,
%``The Early Universe,''
Front.\ Phys.\  {\bf 69} (1990) 1.
%%CITATION = FRPHA,69,1;%%
%1437 citations counted in INSPIRE as of 20 Apr 2018

%\cite{Edsjo:1997bg}
\bibitem{Edsjo:1997bg} 
J.~Edsjo and P.~Gondolo,
%``Neutralino relic density including coannihilations,''
Phys.\ Rev.\ D {\bf 56}, 1879 (1997).
%doi:10.1103/PhysRevD.56.1879
%[hep-ph/9704361].
%%CITATION = doi:10.1103/PhysRevD.56.1879;%%
%548 citations counted in INSPIRE as of 20 Apr 2018

%\cite{Bernal:2017kxu}
\bibitem{Bernal:2017kxu}
N.~Bernal, M.~Heikinheimo, T.~Tenkanen, K.~Tuominen and V.~Vaskonen,
%``The Dawn of FIMP Dark Matter: A Review of Models and Constraints,''
Int.\ J.\ Mod.\ Phys.\ A {\bf 32} (2017) no.27,  1730023.
%doi:10.1142/S0217751X1730023X
%[arXiv:1706.07442 [hep-ph]].
%%CITATION = doi:10.1142/S0217751X1730023X;%%
%40 citations counted in INSPIRE as of 20 Apr 2018



%\cite{Aprile:2012nq}
\bibitem{Aprile:2012nq}
E.~Aprile et al (XENON100 Collaboration),
%``Dark Matter Results from 225 Live Days of XENON100 Data,''
Phys.\ Rev.\ Lett,\  {\bf 109}:181301 (2012).
%doi:10.1103/PhysRevLett.109.181301
%[arXiv:1207.5988 [astro-ph.CO]].


%\cite{Akerib:2013tjd}
\bibitem{Akerib:2013tjd}
D.~S.~Akerib et al (LUX Collaboration),
%``First results from the LUX dark matter experiment at the Sanford Underground Research Facility,''
Phys.\ Rev.\ Lett,\  {\bf 112}:091303 (2014).
%doi:10.1103/PhysRevLett.112.091303
%[arXiv:1310.8214 [astro-ph.CO]].


%\cite{Hochberg:2015pha}
\bibitem{Hochberg:2015pha}
Y.~Hochberg, Y.~Zhao, and K.~M.~Zurek,
%``Superconducting Detectors for Superlight Dark Matter,''
Phys.\ Rev.\ Lett,\  {\bf 116} (1):011301 (2016).
%doi:10.1103/PhysRevLett.116.011301
%[arXiv:1504.07237 [hep-ph]].

\bibitem{Hoch1}
Y.~Hochberg, M.~Pyle, Y.~Zhao et al,
%``Detecting Superlight Dark Matter with Fermi-Degenerate Materials,''
JHEP, {\bf 1608}:057 (2016).
%doi:10.1007/JHEP08(2016)057
%[arXiv:1512.04533 [hep-ph]].

\bibitem{Hoch2}
Y.~Hochberg, Y.~Kahn, M.~Lisanti et al,
%``Directional detection of dark matter with two-dimensional targets,''
Phys.\ Lett.\ B, {\bf 772}:239 (2017).
%doi:10.1016/j.physletb.2017.06.051
%[arXiv:1606.08849 [hep-ph]].


%\cite{Aad:2012tfa}
\bibitem{Aad:2012tfa}
G.~Aad  et al (ATLAS Collaboration),
%``Observation of a new particle in the search for the Standard Model Higgs boson with the ATLAS detector at the LHC,''
Phys.\ Lett.\ B, {\bf 716}:1 (2012).
%doi:10.1016/j.physletb.2012.08.020
%[arXiv:1207.7214 [hep-ex]].

\bibitem{Chatr}
S.~Chatrchyan et al (CMS Collaboration),
%``Observation of a new boson at a mass of 125 GeV with the CMS experiment at the LHC,''
Phys.\ Lett.\ B, {\bf 716}:30 (2012).
%doi:10.1016/j.physletb.2012.08.021
%[arXiv:1207.7235 [hep-ex]].


%\cite{Belanger:2013kya}
\bibitem{Belanger:2013kya}
G.~Belanger, B.~Dumont, U.~Ellwanger et al, 
%``Status of invisible Higgs decays,''
Phys.\ Lett.\ B, {\bf 723}:340 (2013).
%doi:10.1016/j.physletb.2013.05.024
%[arXiv:1302.5694 [hep-ph]].


%\cite{Tulin:2017ara}
\bibitem{Tulin:2017ara}
S.~Tulin, and H.~B.~Yu,
%``Dark Matter Self-interactions and Small Scale Structure,''
arXiv: hepph/1705.02358.
%%CITATION = ARXIV:1705.02358;%%
%48 citations counted in INSPIRE as of 14 Jan 2018

%\cite{Clowe:2003tk}
\bibitem{Clowe:2003tk}
D.~Clowe, A.~Gonzalez, and M.~Markevitch,
%``Weak lensing mass reconstruction of the interacting cluster 1E0657-558: Direct evidence for the existence of dark matter,''
Astrophys.\ J,\  {\bf 604}:596 (2004).
%doi:10.1086/381970
%[astro-ph/0312273].


%\cite{Randall:2007ph}
\bibitem{Randall:2007ph}
S.~W.~Randall, M.~Markevitch, D.~Clowe et al,
%``Constraints on the Self-Interaction Cross-Section of Dark Matter from Numerical Simulations of the Merging Galaxy Cluster 1E 0657-56,''
Astrophys.\ J,\  {\bf 679}:1173 (2008).
%doi:10.1086/587859
%[arXiv:0704.0261 [astro-ph]].


%\cite{Kaplinghat:2015aga}
\bibitem{Kaplinghat:2015aga}
M.~Kaplinghat, S.~Tulin and H.~B.~Yu,
%``Dark Matter Halos as Particle Colliders: Unified Solution to Small-Scale Structure Puzzles from Dwarfs to Clusters,''
Phys.\ Rev.\ Lett.\  {\bf 116} (2016) no.4,  041302.
%doi:10.1103/PhysRevLett.116.041302
%[arXiv:1508.03339 [astro-ph.CO]].
%%CITATION = doi:10.1103/PhysRevLett.116.041302;%%
%100 citations counted in INSPIRE as of 05 Mar 2018



%\cite{Campbell:2015fra}
\bibitem{Campbell:2015fra}
R.~Campbell, S.~Godfrey, H.~E.~Logan et al, 
%``Implications of the observation of dark matter self-interactions for singlet scalar dark matter,''
Phys.\ Rev.\ D, {\bf 92} (5):055031 (2015).
%doi:10.1103/PhysRevD.92.055031
%[arXiv:1505.01793 [hep-ph]].


%\cite{Tulin:2013teo}
\bibitem{Tulin:2013teo}
S.~Tulin, H.~B.~Yu, and K.~M.~Zurek,
%``Beyond Collisionless Dark Matter: Particle Physics Dynamics for Dark Matter Halo Structure,''
Phys.\ Rev.\ D, {\bf 87} (11):115007 (2013).
%doi:10.1103/PhysRevD.87.115007
%[arXiv:1302.3898 [hep-ph]].


%\cite{Kainulainen:2015sva}
\bibitem{Kainulainen:2015sva}
K.~Kainulainen, K.~Tuominen, and V.~Vaskonen,
%``Self-interacting dark matter and cosmology of a light scalar mediator,''
Phys.\ Rev.\ D, {\bf 93} (1):015016 (2016)
Erratum: [Phys.\ Rev.\ D, {\bf 95} (7):079901 (2017)].
%doi:10.1103/PhysRevD.95.079901, 10.1103/PhysRevD.93.015016
%[arXiv:1507.04931 [hep-ph]].



%\cite{Kouvaris:2014uoa}
\bibitem{Kouvaris:2014uoa}
C.~Kouvaris, I.~M.~Shoemaker, and K.~Tuominen,
%``Self-Interacting Dark Matter through the Higgs Portal,''
Phys.\ Rev.\ D, {\bf 91} (4):043519 (2015).
%doi:10.1103/PhysRevD.91.043519
%[arXiv:1411.3730 [hep-ph]].



%\cite{Duch:2017nbe}
\bibitem{Duch:2017nbe}
M.~Duch, and B.~Grzadkowski,
%``Resonance enhancement of dark matter interactions: the case for early kinetic decoupling and velocity dependent resonance width,''
JHEP, {\bf 1709}:159 (2017).
%doi:10.1007/JHEP09(2017)159
%[arXiv:1705.10777 [hep-ph]].


%\cite{Belanger:2018ccd}
\bibitem{Belanger:2018ccd}
G.~Bélanger, F.~Boudjema, A.~Goudelis et al,
%``micrOMEGAs5.0 : freeze-in,''
arXiv: hepph/1801.03509.
%%CITATION = ARXIV:1801.03509;%%


%\cite{Fradette:2014sza}
\bibitem{Fradette:2014sza}
A.~Fradette, M.~Pospelov, J.~Pradler and A.~Ritz,
%``Cosmological Constraints on Very Dark Photons,''
Phys.\ Rev.\ D {\bf 90} (2014) no.3,  035022.
%doi:10.1103/PhysRevD.90.035022
%[arXiv:1407.0993 [hep-ph]].
%%CITATION = doi:10.1103/PhysRevD.90.035022;%%
%45 citations counted in INSPIRE as of 05 Mar 2018

%\cite{Berger:2016vxi}
\bibitem{Berger:2016vxi}
J.~Berger, K.~Jedamzik and D.~G.~E.~Walker,
%``Cosmological Constraints on Decoupled Dark Photons and Dark Higgs,''
JCAP {\bf 1611} (2016) 032.
%doi:10.1088/1475-7516/2016/11/032
%[arXiv:1605.07195 [hep-ph]].
%%CITATION = doi:10.1088/1475-7516/2016/11/032;%%
%14 citations counted in INSPIRE as of 05 Mar 2018

\end{thebibliography}
\end{document}